\documentclass[journal]{IEEEtran}

\usepackage{cite}
\ifCLASSINFOpdf
  \usepackage[pdftex]{graphicx}
\else
  \usepackage[dvips]{graphicx}
\fi

\usepackage[cmex10]{amsmath}
\usepackage{amsmath,amsfonts,amssymb}
\usepackage{mdwmath}

\usepackage{algorithmic}

\usepackage[ruled,lined,boxed]{algorithm2e}

\usepackage{array}
\usepackage{fixltx2e}
\usepackage{bm}
\usepackage{stfloats}
\usepackage{graphicx}
\usepackage{caption}
\usepackage{mdwtab}
\usepackage{subfig}
\usepackage{booktabs}
\usepackage{multirow}
\usepackage{etoolbox}
\usepackage{makecell}
\usepackage{xcolor}
\usepackage[bottom]{footmisc}
\usepackage{mdframed}
\usepackage{steinmetz}

\usepackage{setspace}

\newcommand{\e}{\begin{equation}}
\newcommand{\ee}{\end{equation}}
\newcommand{\eqn}{\begin{eqnarray}}
\newcommand{\eeqn}{\end{eqnarray}}

\begin{document}
\title{Deep Learning-Based Rate-Splitting Multiple Access for Reconfigurable Intelligent Surface-Aided  Tera-Hertz Massive MIMO }

\author{Minghui Wu, Zhen Gao, Yang Huang, Zhenyu Xiao, Derrick~Wing~Kwan~Ng,~\IEEEmembership{Fellow,~IEEE}, and Zhaoyang~Zhang
\vspace*{-5.0mm}
\thanks{M.~Wu and Z.~Gao are with School of Information and Electronics, Beijing Institute of Technology, Beijing 100081, China (e-mails: wuminghui@bit.edu.cn; gaozhen16@bit.edu.cn). Y.~Huang is with the Key Laboratory of Dynamic Cognitive System of Electromagnetic Spectrum Space, Ministry of Industry and Information Technology, Nanjing University of Aeronautics and Astronautics, Nanjing 210016, China (e-mail: yang.huang.ceie@nuaa.edu.cn). Z.~Xiao is with the School of Electronic and Information Engineering, Beihang University, Beijing 100191, China (e-mail: xiaozy@buaa.edu.cn). D.~W.~K.~Ng is with the School of Electrical Engineering and Telecommunications, University of New South Wales, Sydney, NSW 2025, Australia (e-mail: w.k.ng@unsw.edu.au). Z.~Zhang is with College of Information Science and Electronic Engineering, Zhejiang University, Hangzhou 310027, China, and with International Joint Innovation Center, Zhejiang University, Haining 314400, China, and also with Zhejiang Provincial Key Laboratory of Info. Proc., Commun. \& Netw. (IPCAN), Hangzhou 310007, China (e-mail: ning\_ming@zju.edu.cn).}
}

\maketitle

\begin{abstract}
Reconfigurable intelligent surface (RIS) can significantly enhance the service coverage of Tera-Hertz massive multiple-input multiple-output (MIMO) communication systems. However, obtaining accurate high-dimensional channel state information (CSI) with limited pilot and feedback signaling overhead is challenging, severely degrading the performance of conventional spatial division multiple access.
To improve the robustness against CSI imperfection, this paper proposes a deep learning (DL)-based  rate-splitting multiple access (RSMA) scheme for RIS-aided Tera-Hertz multi-user MIMO systems. 
Specifically, we first propose a hybrid data-model driven DL-based RSMA precoding scheme, including the passive precoding at the RIS as well as the analog active precoding and the RSMA digital active precoding at the base station (BS). To realize the passive precoding at the RIS, we propose a Transformer-based data-driven RIS reflecting network (RRN).
As for the analog active precoding at the BS, we propose a match-filter based analog precoding scheme considering that the BS and RIS adopt the LoS-MIMO antenna array architecture.
{\color{black}As for the RSMA digital active precoding at the BS, we propose a  low-complexity approximate weighted minimum mean square error (AWMMSE) digital precoding scheme, and further design a model-driven deep unfolding  active precoding network (DFAPN) by combining the proposed AWMMSE scheme with DL.
Then, to acquire accurate CSI at the BS for the investigated  RSMA precoding scheme to achieve higher spectral efficiency,  we propose  a CSI acquisition network (CAN)  with low pilot and feedback signaling overhead.}
The proposed DL-based RSMA scheme  for RIS-aided Tera-Hertz multi-user MIMO systems can exploit the advantages of RSMA and DL to improve the robustness against CSI imperfection, thus achieving higher spectral efficiency with lower signaling overhead.
\end{abstract}


\IEEEpeerreviewmaketitle

\section{Introduction}\label{S1}
{\color{black}In recent years, Tera-Hertz communications have attracted a great deal of attention from academia and industry due to the significant increase in demand for wireless data traffic \cite{THz,Wan_Tcom}. Compared to the millimeter wave band, the Tera-Hertz band offers abundant unlicensed bandwidth that can achieve higher data rates and lower latency. However, the Tera-Hertz band suffers from strong atmospheric attenuation and free-space loss, and the line-of-sight (LoS) link in the Tera-Hertz band is sensitive to blockages \cite{Wan_Tcom}. These disadvantages severely reduce the coverage of Tera-Hertz communication systems. Deploying massive multiple-input multiple-output (MIMO) in Tera-Hertz communication systems can provide significant beamforming gain and thus increase the coverage range \cite{M_MIMO}. Besides, reconfigurable intelligent surface (RIS) has also been recognized in recent years as a key enabling technology for future wireless communication systems, which can manipulate the amplitude and phase of the incident electromagnetic signals so as to reflect them towards the desired directions and provide beamforming gain \cite{RIS,RIS_zongshu2}. Compared to conventional active relaying, RIS helps to improve the energy efficiency of the system by eliminating the need for power-hungry RF chains and power amplifiers. Therefore, the application of RIS and massive MIMO techniques to Tera-Hertz communications is expected to overcome the above limitations of Tera-Hertz communications.}


In Tera-Hertz band, due to various factors, such as cablibration error of RF chains and the limited uplink transmit power at the UEs for compensating the high path loss, it is challenging to exploit the channel reciprocity to obtain the downlink channel state information (CSI) based on the uplink channel estimation results as in the conventional sub-6 GHz time division duplex systems \cite{THz}.
To this end, in Tera-Hertz massive MIMO-orthogonal frequency division multiplexing (OFDM) systems, the downlink CSI is first estimated at the user equipments (UEs) based on the received downlink pilot signals and then fed back to the BS\cite{THz,FDD}.
As such, deploying a large-scale RIS in Tera-Hertz massive MIMO-OFDM systems can effectively improve the system capacity and energy efficiency \cite{Wan_Tcom}. Yet, it also involves the estimation and feedback of CSI matrices with huge dimensions such that it is challenging  to obtain accurate CSI at the BS with low pilot and feedback signaling overhead. 
On the other hand, conventional space division multiple access (SDMA) techniques rely on the acquisition of  accurate downlink  CSI at the BS \cite{THz}, which suffers from a severe performance loss when the CSI is imperfect. 
Therefore, how to acquire accurate CSI at the BS with low signaling overhead and perform robust precoding in the existence of imperfect CSI is with utmost importance \cite{THz}. 

As a remedy, rate-splitting multiple access (RSMA) has been proposed as a novel and powerful emerging non-orthogonal transmission strategy \cite{RSMA_JSAC,RSMA_zongshu1,RSMA_RIS_zongshu1}.
By splitting the messages into the common and private parts, where the common part is encoded as the common data streams and decoded by multiple UEs while the private part is encoded as the private data streams and decoded by the corresponding UE, RSMA enables a flexible interference management capability and robustness against CSI imperfection \cite{RSMA_Power1,RSMA_Power2,RSMA_Power3}.
In particular, various RSMA precoding techniques have been proposed in recent years to unlock the potential of RSMA \cite{RSMA_DPC,RSMA_Energy,RSMA_Power1,RSMA_Power2,RSMA_Power3,RSMA_SCA,WMMSE_RSMA1,RSMA_RIS1,RSMA_RIS2,RSMA_RIS3}.
Therefore, applying RSMA techniques to the amalgamation of Tera-Hertz massive MIMO and RIS techniques is expected to be promising.


 \subsection{Related Works}\label{S1.1}
 As for the RSMA  precoding, a transmit power allocation-based scheme with imperfect CSI and its derivatives    were respectively proposed in \cite{RSMA_Power1,RSMA_Power2,RSMA_Power3} for different communication systems to enhance robustness. 
Also, the authors of \cite{maxmin,RSMA_SCA,RSMA_DPC,WMMSE_RSMA1} proposed  RSMA precoding schemes with imperfect CSI based on successive convex approximation, dirty paper coding, and weighted minimum mean square error (WMMSE), respectively, to further enhance robustness.
 Besides, an energy efficiency maximization-based RSMA precoding scheme with perfect CSI was proposed in \cite{RSMA_Energy}.
 Furthermore, the authors of \cite{RSMA_RIS1,RSMA_RIS2,RSMA_RIS3} combined the RIS phase optimization with RSMA precoding and proposed RIS-aided RSMA precoding schemes based on alternating optimization with the consideration of perfect CSI.
 These research works demonstrate that RSMA precoding schemes can achieve better performance than conventional SDMA precoding schemes in various practical scenarios.
 However, these existing RSMA precoding schemes either only consider the availability of perfect CSI, e.g., \cite{RSMA_Energy,RSMA_RIS1,RSMA_RIS2,RSMA_RIS3}, or only consider that the CSI errors follow the independent and identically distributed (i.i.d.) zero mean complex Gaussian distribution, e.g., \cite{RSMA_Power1,RSMA_Power2,RSMA_Power3,RSMA_DPC,RSMA_SCA,WMMSE_RSMA1}.
 More importantly, since the practical CSI errors caused by imperfect channel estimation and feedback are more complicated, adopting the above works may lead to severe performance loss.


 Effective precoding in RIS-aided massive MIMO-OFDM systems relies on accurate CSI estimation. To address this issue, the authors of \cite{LS_CE1,LS_CE2} proposed the least square-based CSI eatimation schemes for narrowband and wideband channels, respectively. To reduce the required pilot signaling  overhead, compressed sensing (CS) algorithms have been adopted to exploit the sparsity of the massive MIMO-OFDM channels in both the angular and delay domains \cite{GMMV_SOMP,RIS_AMP,RIS_atomic,RIS_active1,RIS_active2}. 
 For example, one of the existing CS schemes consider that the CSI estimation is performed at the UEs based on the received pilot signals. Specifically, the authors of \cite{GMMV_SOMP,RIS_AMP,RIS_atomic} proposed the CSI estimation schemes based on orthogonal matching tracking (OMP), approximate message passing (AMP), and atomic norm minimization, respectively. However, even with the above CS schemes, estimating high-dimensional CSI at the UEs with a limited number of RF chains still requires a significant pilot signaling overhead. Some other existing schemes consider to equip some active RF chains at the RIS such that CS and deep learning (DL) techniques can be adopted to reconstruct the CSI based on the received pilot signals at the RIS \cite{RIS_active1,RIS_active2}, for further reducing the pilot signaling overhead. However, these solutions \cite{RIS_active1,RIS_active2} require the deployment of expensive power-hungry RF chains at the RIS, which defeats the purpose of reducing hardware cost and power consumption by deploying the passive RIS.

 On the other hand, various feedback schemes have been proposed for massive MIMO-OFDM systems \cite{codebook1,codebook2,codebook33,codebook3,codebook4}.
 One common approach is to adopt CS algorithms to estimate the sparse parameters of the massive MIMO-OFDM channels (e.g. delays and angles of propagation paths) and feed them back to the BS for CSI reconstruction \cite{codebook1,codebook2,codebook33}.
 Also, some other schemes are based on the codebook feedback \cite{codebook3,codebook4}, where the BS or RIS performs a beam scan based on a predefined codebook and the UEs then feed the indices of the selected beams with the highest received signal-to-noise ratios (SNRs) back to the BS for the precoding design.
However, due to the huge CSI dimension of the RIS-aided Tera-Hertz massive MIMO-OFDM channels, the above two schemes still inevitably face excessively high amount of signaling overhead and their high computational complexity as well as the dependency on \emph{a priori} sparsity assumption of channels remain unsolved.

\subsection{Motivations}\label{S1.2}
{\color{black} In recent years, data-driven DL techniques have become a hot research topic in the field of communication physical layer transmission.  }
In terms of channel estimation, the authors of \cite{Ma_tvt} proposed a deep neural network (DNN)-based channel estimation scheme performed at the UEs, while the authors of \cite{RIS_active1,RIS_active2} proposed the DNN-based channel estimation schemes performed at the RIS. Besides, in terms of CSI feedback, a convolutional neural network (CNN)-based scheme, named csiNet, and its derivatives were respectively proposed in \cite{csinet1,csinet2,csinet3} to reduce the feedback signaling overhead and improve the feedback accuracy. {\color{black}As for the precoding design, the authors of \cite{DL_pre1,DL_pre2,DL_pre3} proposed various types of DL-based precoding schemes, and the authors of \cite{RSMA_DL1} utilized deep reinforcement learning technique to solve the power allocation problem in RSMA-based LEO satellite networks.}
These data-driven DL-based schemes do not rely on existing model-based schemes and are capable of learning and optimizing communication transmission strategies directly from training samples.

{\color{black}Besides, model-driven DL techniques have also attracted a lot of attention in the field of communications in recent years. Specifically, model-driven DL is characterized by the construction of neural networks based on the expert knowledge from traditional model-based solutions. Unlike purely data-driven DL solutions, model-driven DL solutions offer better interpretability, predictability, generalization, and faster convergence by introducing expert knowledge. Compared to traditional model-based solutions, model-driven DL solutions introduce DL trainable parameters to mitigate the performance loss caused by model mismatch.} Specifically, \cite{SMV_LAMP} proposed a learnable AMP (LAMP) network by introducing learnable parameters into the conventional AMP algorithm, while \cite{Ma_JSAC} further considered a multicarrier  based multiple-measurement-vectors (MMV) scenario and proposed an MMV-LAMP network. Also, the authors of \cite{DL_unfloding1,DL_unfloding2} unfolded the conventional WMMSE-based precoding algorithm and introduced learnable parameters to achieve better performance. {\color{black}Besides, the authors of \cite{RSMA_DL2} proposed a model-driven DL-based RSMA receiver.}

The DL-based schemes presented above show that both the data-driven and model-driven DL-based schemes are effective in improving the performance of communication systems. Therefore, we consider to combine DL with RSMA, RIS, and Tera-Hertz massive MIMO-OFDM.
{\color{black}Specifically, as for the RSMA digital active precoding for RIS-aided Tera-Hertz massive MIMO-OFDM systems, the model-based approximate WMMSE (AWMMSE) scheme proposed in Section III-D can achieve good performance.
	However, the AWMMSE algorithm faces prohibitively high computational complexity due to the required large number of iterations, and cannot achieve the near-optimal performance caused by the mismatch between the ideal assumptions and the imperfect factors in the practical systems. To this end, we consider to further propose a model-driven DL-based scheme to deep unfold the proposed AWMMSE for achieving better precoding performance and lower computational complexity.}
In terms of channel acquisition, existing CS-based schemes are unable to achieve satisfactory performance with insufficient pilot and feedback signaling overhead. To this end, we propose a purely data-driven DL-based scheme to learn accurate CSI acquisition using low pilot and feedback signaling overhead from training samples.

\subsection{Contributions}\label{S1.3}
This paper proposes a DL-based RSMA precoding scheme and the associated DL-based CSI acquisition scheme for RIS-aided Tera-Hertz massive MIMO-OFDM systems. The main contributions of this paper are summarized as follows:

\begin{itemize}
	\item For passive precoding at the RIS, we propose a RIS reflecting network (RRN) based on the emerging Transformer structure \cite{Transformer,VIT} in DL to design the RIS reflecting matrix. {\color{black}The proposed RRN effectively extracts the correlation features of multiple subcarriers of the MIMO-OFDM channel by using the global feature extraction capability of the Transformer, thus effectively designing a frequency-flat RIS reflecting matrix under a frequency-selective channel.}
\end{itemize}

\begin{itemize}
	\item We derive an AWMMSE-based precoding scheme for the RSMA digital active precoding, which can achieve high achievable rate performance under imperfect CSI wth low computational complexity.
	By combining the derived AWMMSE-based RSMA digital active precoding scheme with DL, we propose a model-driven deep unfolding  active precoding network (DFAPN), where a Transformer is adopted to output the key parameters in the AWMMSE iteration.
	{\color{black}The proposed model-driven DFAPN exploits the \emph{a priori} model in the AWMMSE scheme, while further enhancing performance through DL training, thus providing good performance, interpretability, predictability and fast convergence.}
\end{itemize}

\begin{itemize}
	\item By adopting the achievable rate of the worst UE (ARWU) as the loss function to perform end-to-end (E2E) training of the proposed RRN and DFAPN, the proposed scheme can realize the joint optimization of the passive precoding at the RIS and the RSMA active precoding at the BS for higher ARWU performance over the state-of-the-art schemes.
\end{itemize}


\begin{itemize}
	\item {We introduce the emerging Transformer structure \cite{Transformer,VIT} in DL into the CSI acquisition, thereby proposing an E2E channel acquisition network (CAN). Specifically, the key transmission modules in the channel acquisition including the downlink pilot transmission, uplink CSI feedback at the UEs, and channel reconstruction at the BS are modeled as an E2E neural network based on Transformer. By adopting the normalized mean squared error (NMSE) as the loss function, we can perform data-driven E2E training on the proposed CAN. {\color{black}The proposed CAN utilizes the global feature extraction capability in the Transformer to extract correlation features from multiple subcarriers of the CSI, thus significantly reducing the pilot and feedback overhead.}	}
\end{itemize}

\textit{Notation}: This paper uses lower-case letters for scalars, lower-case bold face letters for column vectors, and upper-case bold face letters for matrices.
Superscripts $(\cdot)^*$, $(\cdot)^T$, $(\cdot)^H$, $(\cdot)^{-1}$, $(\cdot)^\dagger$ denote the conjugate, transpose, conjugate transpose, inversion, and Moore-Penrose inversion operators, respectively.
${\left\| {\mathbf{A}} \right\|_F}$ is the Frobenius norm of ${\mathbf{A}}$, respectively.
${{\rm{vec}}( {\mathbf{A}} )}$ and ${{\rm{angle}}( {\mathbf{A}} )}$ denote the vectorization operation and the phase values of ${\mathbf{A}}$, respectively.
${{\mathbf{I}}_n}$  denotes an identity matrix with size $n \times n$, while $\bm{1}_n$ ($\bm{0}_n$) denotes the vector of size $n$ with all the elements being $1$ ($0$).
${\Re\{\cdot\}}$ and ${\Im\{\cdot\}}$ denote the real part and imaginary part of the corresponding arguments, respectively.
$[\mathbf{A}]_{m,n}$ denotes
the $m$th row and $n$th column element of $\mathbf{A}$, while $[\mathbf{A}]_{[:,m:n]}$
is the sub-matrix containing the $m$th to $n$th columns of $\mathbf{A}$. The expectation is denoted by $\mathbb{E}(\cdot)$. $\frac{{\partial {a }}}{{\partial b}}$ denotes the partial derivative of $a$ with respect to $b$.

\begin{figure*}[t]
	\vspace*{-2mm}
	\centering
	\includegraphics[width = 2 \columnwidth,keepaspectratio]
	{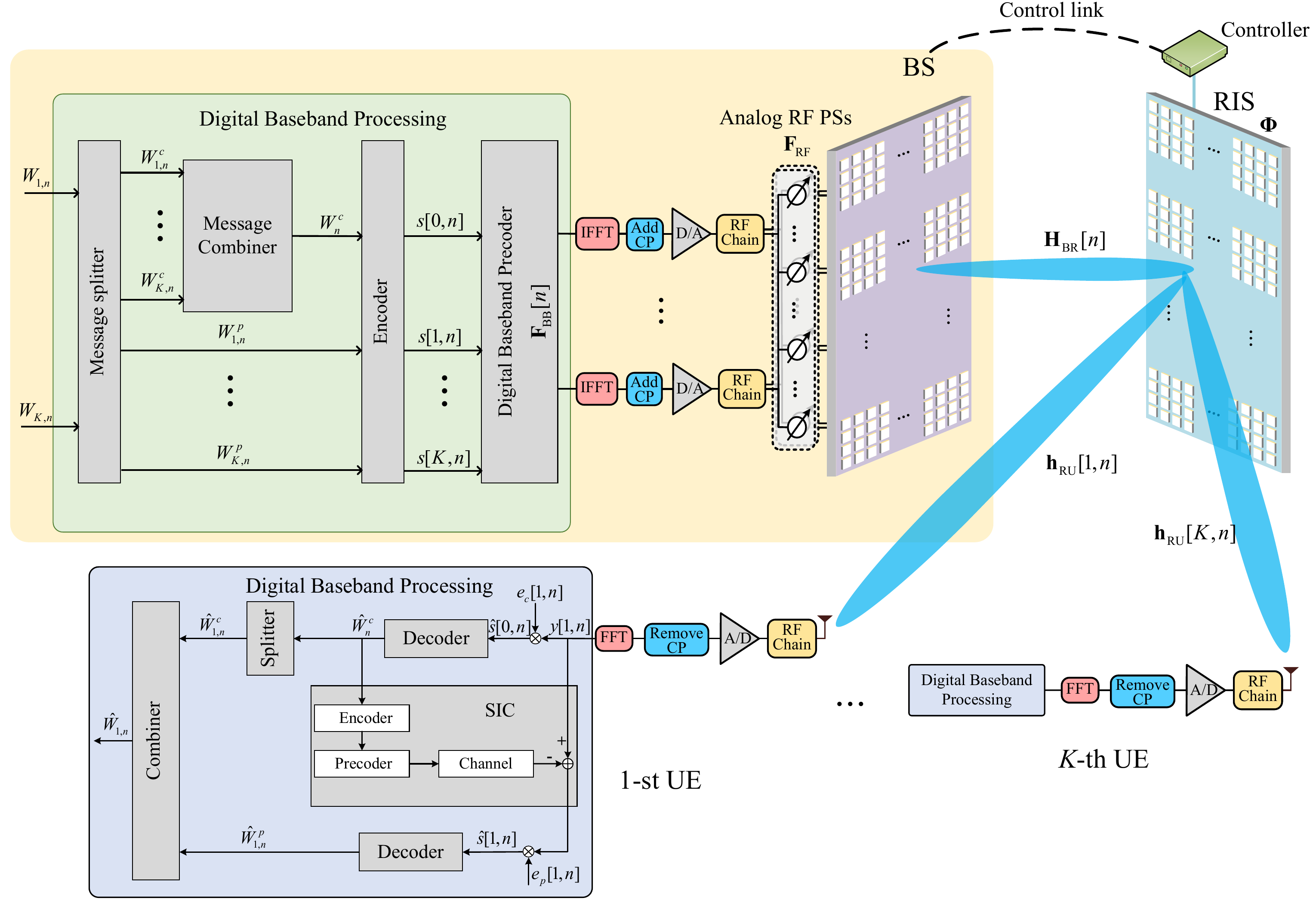}
	\captionsetup{font={footnotesize}, singlelinecheck = off, justification = raggedright,name={Fig.},labelsep=period}
	\caption{RSMA precoding for RIS-aided Tera-Hertz massive MIMO systems.}
	\label{fig:SYS}
	\vspace*{-2mm}
\end{figure*}

\section{System Model}\label{S2}
This paper investigates the downlink multi-user precoding for RIS-aided Tera-Hertz massive MIMO-OFDM systems as shown in Fig. \ref{fig:SYS}, where the LoS paths between the BS and user equipments (UEs) are blocked and the RIS is deployed between the BS and the blocked UEs to establish an E2E virtual LoS link \cite{Wan_Tcom}. Specifically, we consider that the BS adopts a uniform planar array (UPA) with $K$ subarrays as shown in Fig. \ref{fig:SYS}, where the fully-connected hybrid MIMO architecture with $K$ radio frequency (RF) chains and $M_b$ antennas are employed. As for the RIS, we consider that the antenna array structure of the RIS is similar to that of the BS, i.e., $M_r$ reflecting elements are evenly distributed to $K$ subarrays. Furthermore, we consider that the BS simultaneously serves $K$ single-antenna UEs and employs the cyclic prefix (CP)-OFDM with $N_c$ orthogonal subcarriers to combat the frequency selectivity over wideband channels.
\subsection{Signal Transmission Model}
During the transmission, we consider that the message $W_{k,n}$ associated with the $k$-th UE at the $n$-th subcarrier is split into a common part $W_{k,n}^c$ and a private part $W_{k,n}^p$ following the RSMA standard protocol as in \cite{RSMA_DPC,RSMA_Energy,RSMA_Power1,RSMA_Power2,RSMA_Power3,RSMA_SCA,WMMSE_RSMA1,RSMA_RIS1,RSMA_RIS2,RSMA_RIS3}. The common messages of all the UEs are combined and encoded into the common data stream $s_c[{n}]\in\mathbb{C}$, while the private messages are encoded into the private data streams, i.e., $s[{1,n}], \cdots, s[{K,n}]\in\mathbb{C}$. The data streams at the BS ${\mathbf s[n]}=\left[ s_c[{n}], s[{1,n}] \cdots, s[{K,n}] \right]^T \in \mathbb{C}^{(K+1)\times 1}$ are first precoded through the digital baseband precoder ${\mathbf F}_{\rm BB}[n] = \left[ {\mathbf f}_{{\rm BB},c}[n],{\mathbf f}_{\rm BB}[1,n],\cdots,{\mathbf f}_{\rm BB}[K,n] \right] \in \mathbb{C}^{K\times (K+1)}$ and then further processed by the analog RF precoder ${\mathbf F}_{\rm RF}\in\mathbb{C}^{M_b\times K}$. 
After precoding, the signal reaches the RIS through the wireless channel ${\mathbf H}_{\rm BR}[n] \in \mathbb{C}^{M_b \times M_r}$ between the BS and the RIS. Assuming that each reflecting element of the RIS is equipped with an independently adjustable phase that can be controlled by the BS through the control link, then we can model the reflection effect of the RIS on the signal as a diagonal matrix ${\mathbf \Phi}={\rm diag}\left(\left[ \exp\left({{\rm j}\phi_1}\right),\cdots,\exp\left({{\rm j}\phi_{M_r}}\right) \right]\right)\in\mathbb{C}^{M_r \times M_r}${\color{black}\footnote{\color{black}The literature \cite{RIS_nonideal1} considered that the amplitude and the phase shift cannot be adjusted independently. While some advanced materials, such as liquid crystals and graphene, are promising to fabricate the RIS operating at high frequency with wide bandwidths \cite{RIS_nonideal2}. Therefore, we assume that the Tera-Hertz RIS can work at large bandwidths and the amplitude and phase shift can be adjust independently.}}.
Therefore, in the downlink transmission, the signal $y[k,n]\in\mathbb{C}$ received at the $k$-th UE associated with the $n$-th subcarrier can be expressed as\footnote{Since we consider that the direct transmission paths between the BS and the UEs are blocked, the BS can only establish indirect transmission paths with the UEs through the RIS. Furthermore, we consider the widely adopted independent diffusive scatterer-based RIS model \cite{RIS_zongshu2}, in which each element of the RIS is treated as an independent scatterer and the reflection effect of the RIS on the wireless signals is modeled as a diagonal matrix satisfying the unit modulus constraint.}
\begin{equation}
	y[k,n]= {\mathbf h}_{\rm RU}^H[k,n]{\mathbf \Phi}{\mathbf H}_{\rm BR}^H[n]{\mathbf F}_{\rm RF}{\mathbf F}_{\rm BB}[n]{\mathbf s}[{n}] + z[k,n],
\end{equation}
where ${\mathbf h}_{\rm RU}[k,n]\in\mathbb{C}^{M_r\times 1}$ is the downlink channel vector between the RIS and the $k$-th UE and ${z}[{k,n}]\sim {\cal CN}\left( {0},\sigma_n^2 \right)$  is the additive white Gaussian noise (AWGN). Furthermore, three precoding constraints are considered in this paper: {\color{black}(1) a power constraint, where the digital precoder should satisfy $\|{\mathbf F}_{\rm RF}{\mathbf F}_{\rm BB}[n]\|_F^2\le P_t,\forall n$, and $P_t$ is the maximum transmit power;} (2) a unit modulus constraint for the BS analog precoder ${\mathbf F}_{\rm RF}\in\cal{F_{\rm RF}}$, where ${\cal{F_{\rm RF}}}=\{{\mathbf F}_{\rm RF}\in{\mathbb{C}}^{M_b\times K} | \left| [{\mathbf F}_{\rm RF}]_{i,j} \right|=1/\sqrt{M_b},\forall i,j \}$; (3) a unit modulus constraint for the RIS reflecting matrix ${\mathbf \Phi}\in\cal{F_{\rm RIS}}$, where ${\cal{F_{\rm RIS}}}=\{{\mathbf \Phi}\in{\mathbb{C}}^{M_r\times M_r} | \left| [{\mathbf \Phi}]_{i,j} \right|=1,{\rm if \ }i=j, \left| [{\mathbf \Phi}]_{i,j} \right|=0, {\rm others}\}$.

At the UEs, each UE first decodes the common data stream by treating all the private data streams as interference \cite{WMMSE_RSMA1}. {\color{black}Then the received common data stream and the corresponding interference and noise can be expressed as ${\bf{h}}_{{\rm{RU}}}^H[k,n]{\bf{\Phi H}}_{{\rm{BR}}}^H[n]{{\bf{F}}_{{\rm{RF}}}}{{\bf{f}}_{{\rm{BB,c}}}}[n]$ and $\sum\limits_{i = 1}^K {\bf{h}}_{{\rm{RU}}}^H[k,n]{\bf{\Phi H}}_{{\rm{BR}}}^H[n]$ ${{\bf{F}}_{{\rm{RF}}}}{{\bf{f}}_{{\rm{BB}}}}[i,n]  + z[k,n]$, respectively. Based on the above definition, the signal-to-interference-plus-noise ratio (SINR) of the common data stream $s_c[{n}]$ at the $n$-th subcarrier associated with the $k$-th UE is given by}
\begin{equation}
	{\rm{SINR}}_{k,n}^c = \frac{{{{\left| {{\bf{h}}_{{\rm{RU}}}^H[k,n]{\bf{\Phi H}}_{{\rm{BR}}}^H[n]{{\bf{F}}_{{\rm{RF}}}}{{\bf{f}}_{{\rm BB},c}}[n]} \right|}^2}}}{{\sum\limits_{i = 1}^K {{{\left| {{\bf{h}}_{{\rm{RU}}}^H[k,n]{\bf{\Phi H}}_{{\rm{BR}}}^H[n]{{\bf{F}}_{{\rm{RF}}}}{{\bf{f}}_{{\rm{BB}}}}[i,n]} \right|}^2} + \sigma _n^2} }},
\end{equation}
and its corresponding achievable rate can be obtained as $R_{k,n}^c={\rm log}_{2}\left( 1+ 	{\rm{SINR}}_{k,n}^c \right)$. To ensure that the common data stream $s_c[n]$ can be decoded by all the UEs, the achievable rate of the common data stream should satisfy $R_{n}^c={\min}_{k}\left\{R_{k,n}^c\right\}$.
As $R_{n}^c$ is shared by all the UEs, it should staisfy $R_n^c = \sum\nolimits_{k = 1}^K {{C_{k,n}}}$, where $C_{k,n}$ is the portion of $R_n^c$ transmitting $W_{k,n}^c$ \cite{maxmin}. 
To simplify the system model, we consider that $R_{n}^c$ is evenly allocated to each UE, i.e., $C_{k,n} = R_{n}^c/K$.
{\color{black}After decoding the common data stream, the $k$-th UE applies successive interference cancellation (SIC) to remove the common data stream from the received signal \cite{WMMSE_RSMA1}, i.e., ${y_p}[k,n] = y[k,n] - {\bf{h}}_{{\rm{RU}}}^H[k,n]{\bf{\Phi H}}_{{\rm{BR}}}^H[n]{{\bf{F}}_{{\rm{RF}}}}{{\bf{f}}_{{\rm{BB,c}}}}[n]{s_c}[n]$. 
Then the received private data stream and the corresponding interference and noise can be expressed as ${\bf{h}}_{{\rm{RU}}}^H[k,n]{\bf{\Phi H}}_{{\rm{BR}}}^H[n]{{\bf{F}}_{{\rm{RF}}}}{{\bf{f}}_{{\rm{BB}}}}[k,n]s[k,n]$ and $\sum\limits_{i = 1,i \ne k}^K {{\bf{h}}_{{\rm{RU}}}^H[k,n]{\bf{\Phi H}}_{{\rm{BR}}}^H[n]{{\bf{F}}_{{\rm{RF}}}}{{\bf{f}}_{{\rm{BB}}}}[i,n]s[k,n] + z[k,n]}$, respectively.
Based on this, the private data stream can be decoded with SINR}
\begin{equation}
	{\rm{SINR}}_{k,n}^p = \frac{{{{\left| {{\bf{h}}_{{\rm{RU}}}^H[k,n]{\bf{\Phi H}}_{{\rm{BR}}}^H[n]{{\bf{F}}_{{\rm{RF}}}}{{\bf{f}}_{{\rm{BB}}}}[k,n]} \right|}^2}}}{{\sum\limits_{i = 1,i\neq k}^K {{{\left| {{\bf{h}}_{{\rm{RU}}}^H[k,n]{\bf{\Phi H}}_{{\rm{BR}}}^H[n]{{\bf{F}}_{{\rm{RF}}}}{{\bf{f}}_{{\rm{BB}}}}[i,n]} \right|}^2} + \sigma _n^2} }},
\end{equation}
and its corresponding achievable rate can be obtained as $R_{k,n}^p={\rm log}_{2}\left( 1+ 	{\rm{SINR}}_{k,n}^p \right)$. 
Then, the achievable rate of the $k$-th user at the $n$-th subcarrier is given by
\begin{equation} \label{equ:SE}
	{R_{k,n}} =   {R_{k,n}^p + C_{k,n}} =  R_{k,n}^p + {\min}_{k}\left\{ R_{k,n}^c\right\}/K.
\end{equation}

\subsection{Channel Model}
Existing studies have exploited non-LoS (NLoS) paths in MIMO channels for additional spatial multiplexing gain \cite{Sun_TWC,OMP_HP}. 
However, for the Tera-Hertz massive MIMO channels considered in this paper, the path loss of the NLoS paths is significantly severer than that for the LoS path due to the ultra-high carrier frequency \cite{Wan_Tcom}. Thus, we consider the massive MIMO channels of BS-RIS and RIS-UE are LoS dominated.
On the other hand, the LoS communication results in low-rank massive MIMO channels such that it is challenging to adopt one RIS to reflect multi-stream signals for serving multiple UEs.
To this end, we consider adopting the LoS-MIMO antenna array architecture at the BS and RIS introduced in \cite{LoS_MIMO}, which preserves the multi-rank of the MIMO channel through specifically designed antenna arrays, thereby offering additional spatial multiplexing gain even over LoS dominated channels. 
{\color{black}Specifically, we consider that both the BS and RIS are equipped with a planar array on the $yz$-plane and point toward each other with a distance $T$, and that both the BS and RIS have $K_y$ and $K_z$ subarrays uniformly distributed on the $y$ and $z$ axes. Then, the subarray spacing of the BS and RIS needs to meet  ${D_y} = \sqrt {{{{\lambda _c}T}}/{{{K_y}}}}$ and ${D_z} = \sqrt {{{{\lambda _c}T}}/{{{K_z}}}}$, so as to ensure the orthogonality between the subarrays.}
The specific antenna array design parameters are given in Section V and the detailed derivation can be found in \cite{LoS_MIMO}.

In this paper, we consider the channel model in \cite{LoS_MIMO}. 
Specifically,  for the mathematical representation of the wireless channel between the BS and the RIS, the channel gain between the $i$-th antenna element of the BS and the $j$-th reflecting element of the RIS (i.e., the element of the $i$-th row and the $j$-th column of the matrix ${\mathbf H}_{\rm BR}[n]$) can be expressed as\footnote{\color{black}Beam squint is an important effect for broadband massive MIMO-OFDM systems \cite{beam_squint}. In this paper, we do not simply use the steering vector to model the channel, but instead calculate the corresponding channel response values based on the frequency of each subcarrier and the coordinates of each array element. This channel modeling approach takes into account not only the additional frequency-selective effect of beam squint (without neglecting the path delay difference of each element) but also the effect of near field (without neglecting the angle difference of each element).  However, the beam squint effect is almost negligible for a single subarray since we use the subarray MIMO architecture and the size of each subarray is very small. To this end, we do not emphasize the impact of beam squint in this paper.}
{\color{black}\begin{equation}\label{equ:h_BR}
	{H_{{\rm{BR}}}}[n,i,j] = \sqrt{{G_{{\rm{BR}}}}[n,i,j]G_t}\exp (\dfrac{{ - {\rm{j}}2\pi ({d_{\rm BR}[{i,j}]})}}{{{\lambda _n}}}),
\end{equation}
where ${G_{{\rm{BR}}}}[n,i,j] ={\lambda_n^2}/({{16\pi^2 d_{\rm BR}^2[{i,j}]}})$ represents the 
channel gain with respect to large-scale fading, $G_t$ = 17 dB is the antenna gain at the BS\footnote{\color{black}The channel fading in the Tera-Hertz band is severe, thus directional antennas need to be used to offer an additional antenna gain, which is typically between 10-20 dB \cite{Gain}. In this paper, we set both the transmit antenna gain of the BS and the receive antenna gain of the UE to 17 dB.}, $\lambda_n$ is the wavelength associated with the $n$-th subcarrier, and $d_{\rm BR}[{i,j}]$ is the distance between the $i$-th antenna element of the BS and the $j$-th reflecting element of the RIS. }

Similarly, the channel gain between the $j$-th reflecting element of the RIS and the $k$-th UE (i.e., the $j$-th element of the vector ${\mathbf h}_{\rm RU}[k,n]$) can be expressed as
{\color{black}
\begin{equation} \label{equ:h_RU}
	h_{\rm RU}[k,n,j] = \sqrt{{G_{{\rm{RU}}}}[k,n,j]G_r}\exp (\dfrac{{ - {\rm{j}}2\pi ({d_{\rm RU}[{j,k}]})}}{{{\lambda _n}}}),
\end{equation}
where  ${G_{{\rm{RU}}}}[k,n,j] = {\lambda_n^2}/({{16\pi^2 {d_{\rm RU}^2[{j,k}]}}})$ is the large-scale fading gain, $G_r$ = 17 dB is the antenna gain at the UE, and $d_{\rm RU}[{j,k}]$ is the distance between the $j$-th reflecting element of the RIS and the $k$-th UE. }

\section{Proposed DL-Based Hybrid Data-Model Driven RSMA Precoding Scheme} \label{S:precoding}

\begin{figure*}[t]
	\vspace*{-3mm}
	\centering
	\includegraphics[width = 2 \columnwidth,keepaspectratio]
	{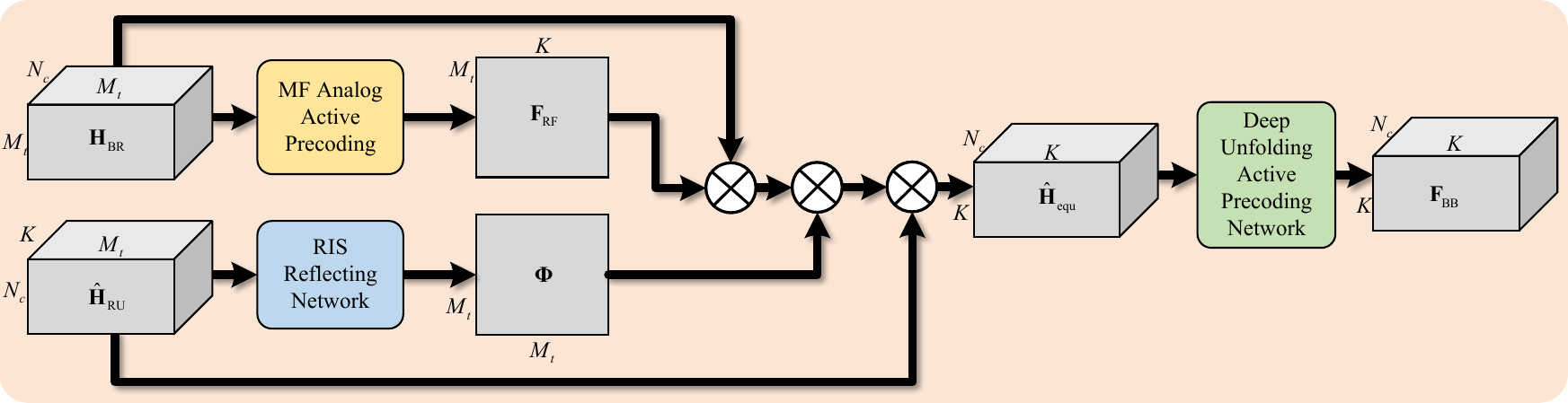}
	\captionsetup{font={footnotesize}, singlelinecheck = off, justification = raggedright,name={Fig.},labelsep=period}
	\caption{The block diagram of the proposed DL-based precoding scheme.}
	\label{fig:precoding}
	\vspace*{-1mm}
\end{figure*}

In this section, we first present the processing procedure of the proposed scheme to design the RSMA precoding for RIS-aided Tera-Hertz massive MIMO systems. Secondly, we design a matched-filter (MF)-based analog precoding scheme for LoS-MIMO antenna array architecture. Thirdly, we propose a Transformer-based RRN to design the RIS reflecting matrix. Fourthly, for the RSMA digital active precoding, we derive a low-complexity AWMMSE algorithm and further propose the DFAPN by deep unfolding the proposed AWMMSE for better precoding performance and lower computational complexity.
 Finally, the proposed RRN and DFAPN are jointly trained to achieve high spectral efficiency taking into the impact of imperfect CSI.
The block diagram of the proposed scheme is shown in Fig. \ref{fig:precoding}.

\subsection{Processing Procedure and Problem Formulation for the RSMA Precoding}
We assume that the positions of the BS and the RIS are fixed such that the downlink BS-RIS CSI is static which can be estimated accurately \cite{Wan_Tcom}.
Therefore, we only consider the estimation of the downlink RIS-UE CSI at the BS, and the BS needs to perform precoding based on the estimated imperfect downlink RIS-UE CSI, i.e.,
\begin{align}
	\left\{ {{\bf{\Phi }},{{\bf{F}}_{{\rm{RF}}}},{{\bf{F}}_{{\rm{BB}}}}[1], \cdots ,{{\bf{F}}_{{\rm{BB}}}}[{N_c}]} \right\} = \nonumber \\  {\cal P}({\widehat{\bf{H}}_{{\rm{RU}}}}[1],\cdots,{\widehat{\bf{H}}_{{\rm{RU}}}}[K]),
\end{align}
where $\widehat{\mathbf H}_{\rm RU}[k]=[\hat{\mathbf h}_{\rm RU}[k,1],\cdots,\hat{\mathbf h}_{\rm RU}[k,N_c]]^H\in\mathbb{C}^{N_c\times M_r}$ is the estimated values of the downlink RIS-UE CSI ${\mathbf H}_{\rm RU}[k]=[{\mathbf h}_{\rm RU}[k,1],\cdots,{\mathbf h}_{\rm RU}[k,N_c]]^H\in\mathbb{C}^{N_c\times M_r}$
 and ${\cal P}(\cdot)$ represents a mapping function where the RIS reflecting matrix, analog precoder, and digital precoder can be a function of the estimated downlink RIS-UE CSI.

Based on (\ref{equ:SE}), the ARWU at the $n$-th subcarrier can be expressed as
\begin{align}
	{R^w_n} = {{\min} _k}\left\{{R_{k,n}}\right\} =  {{\min}_k } \left\{{R_{k,n}^p}\right\}   + {{\min}_k }\left\{{ {{}R_{k,n}^c} }\right\}.
\end{align}

We propose to maximize the ARWU, then the design of precoding can be formulated as the following optimization problem, i.e.,
\begin{equation}
	\label{equ:P2}
\begin{array}{*{20}{l}}
	\begin{array}{l}
		\mathop {{\rm{maximize}}}\limits_{{\cal P}( \cdot )} \quad \\
		\\
		
	\end{array}&\begin{array}{l}
		{R^w} = \frac{1}{{{N_c}}}\sum\limits_{n = 1}^{{N_c}} {R_n^w}=\frac{1}{{{N_c}}} \\
		{\rm{     }}  \sum\limits_{n = 1}^{{N_c}} {\left( {{{\min }_k}\left\{ {R_{k,n}^p} \right\} + {{\min }_k}\left\{ {R_{k,n}^c} \right\}} \right)}, 
	\end{array}\\
	\quad\quad\quad{{\rm{s}}.{\rm{t}}.\quad }&{\left\{ {{\bf{\Phi }},{{\bf{F}}_{{\rm{RF}}}},{{\bf{F}}_{{\rm{BB}}}}[1], \cdots ,{{\bf{F}}_{{\rm{BB}}}}[{N_c}]} \right\}}\\
	{}&\begin{array}{l}
		{\rm{            }} = {\cal P}({\widehat {\bf{H}}_{{\rm{RU}}}}[1], \cdots ,{\widehat {\bf{H}}_{{\rm{RU}}}}[K]),{\rm{                 }}
	\end{array}\\
    & {\mathbf \Phi}\in {\cal{F_{\rm RIS}}},  \nonumber \\
	{}&{{{\bf{F}}_{{\rm{RF}}}} \in {F_{{\rm{RF}}}},}\\
	{}&{{{\bf{F}}_{{\rm{RF}}}}{{\bf{F}}_{{\rm{BB}}}}[n]_F^2 \le {P_t},\forall n.}
\end{array}
\end{equation}

In the following, we present the schemes proposed for the above optimization problem.


\subsection{MF-Based Analog Active Precoding at the BS} \label{s:MF}
Given the fact that the BS-RIS channel is known and LoS-dominated and that the BS and RIS adopt the LoS-MIMO antenna array architecture \cite{LoS_MIMO}, we adopt a simple MF strategy to design the analog precoder to maximize the capacity of the LoS-dominated BS-RIS link, which is briefly introduced below.

Specifically, we consider the normal directions of the planar arrays of the BS and the RIS point towards each other and the subarray apertures are small, as shown in Fig. \ref{fig:SYS}. To this end, the channel phase difference of the antenna array elements within a subarray can be ignored, and only the channel phase   between subarrays $\theta [{k_1},{k_2}]$ can be considered, i.e.,
\begin{align}
	\theta [{k_1},{k_2}] = {\rm angle} & \left({\left[ {{{\bf{H}}_{{\rm{BR}}}}[N_c/2]} \right]_{[({k_1} - 0.5){M_{\rm sub}^b},({k_2} - 0.5){M_{\rm sub}^r}]}}\right),\nonumber\\ & 1 \le {k_1},{k_2} \le K,
\end{align}
where  $M_{\rm sub}^b=M_b/K$ and $M_{\rm sub}^r=M_r/K$ are the number of elements in each subarray of the BS and RIS, respectively, $k_1$ and $k_2$ indicate the subarray indexes at the BS and RIS, respectively. Besides, we consider to use the channel phase of the central element of each subarray at the central subcarrier to approximate the channel phase of the whole subarray. 
Then, the CSI between the $k_1$-th subarray at the BS and the $k_2$-th subarray at the RIS can be approximated as\footnote{Note that the approximation here is for analog precoding only, thus the factors such as large-scale fading, frequency-selective fading due to time delay differences between subarrays, etc. are ignored. }
\begin{align}
{{\bf{H}}_{{\rm{sub}}}}[{k_1},{k_2}] = \exp ({\rm{j}}\theta [{k_1},{k_2}]){{\bf{1}}_{{M_{{\rm{sub}}}^b}}}{\bf{1}}_{{M_{{\rm{sub}}}^r}}^H \in \mathbb{C} {^{{M_{{\rm{sub}}}^b} \times {M_{{\rm{sub}}}^r}}}.
\end{align}

Since we adopt the LoS-MIMO architecture, the channels between different subarrays can be considered to be approximately orthogonal to each other, i.e.,
\begin{align}
	\left\{ {\begin{array}{*{20}{l}}
			{\sum\limits_{{k_1}}^K {\exp ( - {\rm{j}}\theta [{k_1},k_2^{(1)}])\exp ({\rm{j}}\theta [{k_1},k_2^{(2)}])}  \approx K, {\rm{ if }}\ k_2^{(1)} = k_2^{(2)},}\\
			{\sum\limits_{{k_1}}^K {\exp ( - {\rm{j}}\theta [{k_1},k_2^{(1)}])\exp ({\rm{j}}\theta [{k_1},k_2^{(2)}])}   \approx  0, {\rm{ others.}}}
	\end{array}} \right.
\end{align}

Therefore, we adopt the MF strategy that allows each analog precoding vector to match the phase of the corresponding subarray to eliminate inter-subarray interference and keep the channel rank unchanged after analog precoding, thus maximizing the capacity of the BS-RIS link and ensuring that multiple UEs can be served simultaneously via the RIS, i.e.,
\begin{align}
	{\left[ {{{\bf{F}}_{{\rm{RF}}}}} \right]_{{(k_1-1)}{M_{\rm sub}^b} + 1:{k_1}{M_{\rm sub}^b},{k_2}}} = & \exp ({\rm{j}}\theta [{k_1},{k_2}]){{\bf{1}}_{{M_{{\rm{sub}}}^b}}}/\sqrt {{M_b}} ,\nonumber\\ & {\rm{ }}1 \le {k_1},{k_2} \le K.
\end{align}
The detailed derivation can be found in \cite{LoS_MIMO}.

\subsection{Proposed RRN for the Passive Precoding at the RIS}
\begin{figure*}[t]
	\vspace*{-3mm}
	\centering
	\includegraphics[width = 2 \columnwidth,keepaspectratio]
	{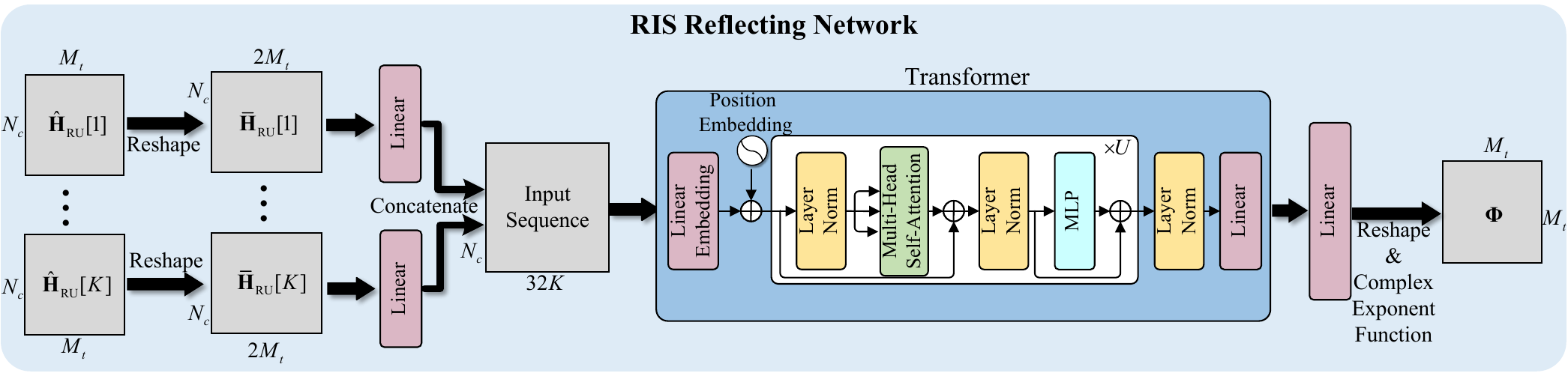}
	\captionsetup{font={footnotesize}, singlelinecheck = off, justification = raggedright,name={Fig.},labelsep=period}
	\vspace{-1mm}
	\caption{The proposed RRN for the design of the RIS reflecting matrix.}
	\label{fig:RRN}
	\vspace*{-1mm}
\end{figure*}


Based on the estimated RIS-UE CSI of all the UEs, the BS needs to design the RIS reflecting matrix
to establish the BS-UE virtual LoS link and improve the system capacity, which can be achieved by the proposed Transformer-based RRN as shown in Fig. \ref{fig:RRN}.
Note that the Transformer structure  in DL has been widely applied in the field of natural language processing and computer vision \cite{Transformer,VIT},  and has been shown to outperform fully connected neural network, CNN, etc. in many scenarios.
{\color{black}Compared with the convolution in CNN \cite{csinet1}, which can only extract features from local areas, self-attention in Transformer can extract global features. Therefore, it can globally extract the inter-subcarrier correlation of the input signal and provide the corresponding weighting coefficients of the components in each subcarrier for enhanced performance.}
The standard Transformer takes 1D real-valued sequence as the input and also takes 1D real-valued sequence as the output \cite{Transformer,VIT}.
To process the complex-valued input ${\widehat {\bf{H}}_{{\rm{RU}}}}[k]$, the RRN first converts it into a real-valued matrix ${{{{\bf{\bar H}}}_{{\rm{RU}}}}[k]}\in\mathbb{R}^{N_c\times 2M_r}$, i.e.,
\begin{align}
\left\{ {\begin{array}{*{20}{l}}
	\ \ \ \ \ 	{{{\left[ {{{{\bf{\bar H}}}_{{\rm{RU}}}}[k]} \right]}_{[:,1:{M_r}]}} = \Re \left\{ {{{{\bf{\widehat H}}}_{{\rm{RU}}}}[k]} \right\}},\\
		{{{\left[ {{{{\bf{\bar H}}}_{{\rm{RU}}}}[k]} \right]}_{[:,1 + {M_r}:2{M_r}]}} = \Im \left\{ {{{{\bf{\widehat H}}}_{{\rm{RU}}}}[k]} \right\}}.
\end{array}} \right.
\end{align}
As shown in Fig. \ref{fig:RRN}, the real-valued CSI matrices of all the UEs are first processed by a fully-connected linear layer to compress their dimensions to $N_c\times 32$ and then concatenated into a 1D real-valued sequence with dimension $N_c \times 32K$ as the input of Transformer, where the number $N_c$ of subcarriers serves as the effective input sequence length for the Transformer.
In the Transformer, the input sequence is first converted into a sequence of vectors with dimension $d_{\rm model}$ by using a fully-connected linear embedding layer and the following position embedding layer, where a sine function of different frequencies is used to represent the positions of different subcarriers.
Then, the Transformer further employs $U$ identical layers to extract the features of the input sequence. Each layer consists of a multi-head self-attention sublayer and a multilayer perceptron (MLP) sublayer.
 After that, the extracted features are processed by a fully-connected linear layer to output the phase values ${\mathbf \Theta}_{\rm RIS}\in\mathbb{R}^{M_r\times M_r}$ of the RIS reflecting matrix.
By applying the complex exponential function to the phase matrix ${\mathbf \Theta}_{\rm RIS}$, the RRN can generate a diagonal RIS reflecting matrix satisfying the unit modulus constraint $\cal{F_{\rm RIS}}$, i.e.,
\begin{align}
	\left\{ {\begin{array}{*{20}{l}}
			{{{\left[ {{\bf{\Phi }}} \right]}_{i,j}} = \exp (1{\rm{j}} \cdot {{\left[ {{\bf{\Theta }}_{{\rm{RIS}}}} \right]}_{i,j}}),\ \ {\rm{  if }}\  i = j,}\\
			{{{\left[ {{\bf{\Phi }}} \right]}_{i,j}} = 0,\ \ {\rm{  others}}}.
	\end{array}} \right.
\end{align}

The above RIS reflecting matrix design process can be expressed as
\begin{align}
	{\mathbf \Phi} ={\cal A}(\widehat{\mathbf H}_{\rm RU}[1],\cdots,\widehat{\mathbf H}_{\rm RU}[K];{\mathcal W}_{\rm RR}),
\end{align}
where ${\cal A}(\cdot;{\mathcal W}_{\rm RR})$ is the mapping function from all the $K$ UEs'  reconstructed downlink RIS-UE CSI to the RIS reflecting matrix and ${\mathcal W}_{\rm RR}$ is the learnable neural network parameters.

\subsection{Proposed AWMMSE-Based RSMA Digital Active Precoding at the BS} \label{SS:WMMSE}
In this subsection, we propose a low-complexity AWMMSE algorithm for the design of RSMA digital active precoding.
With the known BS-RIS CSI ${{\bf{H}}_{{\rm{BR}}}}[n]$, the designed RIS reflecting matrix ${{\bf{\Phi }}}$ as well as the analog active precoding matrix ${{\bf{F}}_{{\rm{RF}}}}$, and the estimated RIS-UE CSI ${{\bf{\hat h}}_{{\rm{RU}}}}[k,n]$ (obtained in Section \ref{S:CSI_acu}), the BS can obtain the estimated equivalent baseband CSI
\begin{align}
	{{\bf{\hat h}}_{{\rm{equ}}}}[k,n] = {{\bf{F}}_{{\rm{RF}}}^H}{{\bf{H}}_{{\rm{BR}}}}[n]{{\bf{\Phi }}^H}{{\bf{\hat h}}_{{\rm{RU}}}}[k,n]\in\mathbb{C}^{K\times 1},\forall k,n.
\end{align}
To this end, the optimization problem in (\ref{equ:P2}) can be simplified as {\color{black}
\begin{align}
	\mathop{\rm maximize}\limits_{{{\bf{F}}_{{\rm{BB}}}}[1], \cdots ,{{\bf{F}}_{{\rm{BB}}}}[{N_c}]}\quad &      
	\frac{1}{N_c}\sum\limits_{n = 1}^{{N_c}}\left(   {{{{\min}_k }}\left\{ {R_{k,n}^p}\right\} }  +  {{{{\min}_k }}\left\{R_{k,n}^c\right\}}   \right) \nonumber \\
	{\rm s.t.} \quad 
	& \|{\mathbf F}_{\rm RF}{\mathbf F}_{\rm BB}[n]\|_F^2\le P_t,\forall n.
\end{align}}
Clearly, the digital active precoding can be designed independently at different subcarriers.  To simplify the mathematical expression, in this subsection, we consider the optimization problem at one of the subcarriers and discard the subcarrier index $n$, that yields,{\color{black}
\begin{align}
	\mathop{\rm maximize}\limits_{{{\bf{F}}_{{\rm{BB}}}}}\quad &   {R^w} =    
	\left(   {{{{\min}_k }}\left\{ {R_{k}^p} \right\}}  +  {{{{\min}_k }}\left\{R_{k}^c\right\}}   \right), \nonumber \\
	{\rm s.t.} \quad 
	& \|{\mathbf F}_{\rm RF}{\mathbf F}_{\rm BB}\|_F^2\le P_t.
\end{align}}
Let $\hat{s}_c[k] = e_c[k]y[k]$ and $\hat{s}[k] = e[k] ({y[k]-\hat{\mathbf h}_{\rm equ}^H[k] {\mathbf f}_{{\rm BB},c}{s}_c})$ be the $k$-th UE's estimates of the common data stream $s_c$ and the private data stream $s[k]$ respectively, where $e_c[k]$ and $e[k]$ are the corresponding equalizers. Then, the 
mean square errors (MSEs) of decoding the common and private data streams can be approximated as\footnote{{\color{black}Since the perfect CSI is not available at the BS during the precoding stage, the MSEs are random variables for the BS that does not facilitate the calculation of the actual achievable rates \cite{WMMSE_RSMA1}. Therefore, we consider perform precoding optimization at the BS using approximate MSEs and rates. In this paper, we assume that the equivalent CSI is available at the UE-side, such that the ARWU considered is achievable and the MSEs come exclusively from inter-stream interference and noise. In the digital precoding stage, we assume that the baseband CSI can be expressed ${{\bf{h}}_{{\rm{equ}}}}[k] = {{\bf{\hat h}}_{{\rm{equ}}}}[k] + {{\bf{\tilde h}}_{{\rm{equ}}}}[k], \forall k$, where ${{\bf{\tilde h}}_{{\rm{equ}}}}[k]$ is the CSI error at the BS. Assume that the UE uses $e_c[k]$ and $e[k]$ as equalisers for the common and private data streams, respectively. For the common data stream, its expected received signal is ${e_c}[k]\left( {{{{\bf{\tilde h}}}_{{\rm{equ}}}}[k] + {{{\bf{\hat h}}}_{{\rm{equ}}}}[k]} \right){{\bf{f}}_{{\rm{BB}},{\rm{c}}}}{s_c}$ and the inter-stream interference and noise is $\sum\nolimits_{i = 1}^K {{e_c}[k]\left( {{{{\bf{\tilde h}}}_{{\rm{equ}}}}[k] + {{{\bf{\hat h}}}_{{\rm{equ}}}}[k]} \right){{\bf{f}}_{{\rm{BB}}}}[i]s[i]}  + {e_c}[k]z[k]$. For the private data stream, its expected received signal is $e[k]\left( {{{{\bf{\tilde h}}}_{{\rm{equ}}}}[k] + {{{\bf{\hat h}}}_{{\rm{equ}}}}[k]} \right){{\bf{f}}_{{\rm{BB}}}}[k]s[k]$ and the inter-stream interference and noise is $\sum\nolimits_{i = 1,i \ne k}^K {e[k]\left( {{{{\bf{\tilde h}}}_{{\rm{equ}}}}[k] + {{{\bf{\hat h}}}_{{\rm{equ}}}}[k]} \right){{\bf{f}}_{{\rm{BB}}}}[i]s[i]}  + e[k]z[k]$. Based on the above conclusions, the MSEs of the common and private data streams can be approximated as formula (\ref{equ:eps}). Note that we simply assume that the CSI error ${{{{\bf{\tilde h}}}_{{\rm{equ}}}}[k]}$ is i.i.d. complex Gaussian distributed and independent of ${{{{\bf{ h}}}_{{\rm{equ}}}}[k]}$, thus the approximate sign is used.}}
\begin{align}
	\label{equ:eps}
	\begin{array}{l}
		{\varepsilon _c}[k]  = \mathbb{E}\left\{ {{{\left| {{{\hat s}_c}[k] - s_c} \right|}^2}} \right\} \approx \varepsilon _c^{(1)}[k] + \varepsilon _c^{(2)}[k],\\
		\varepsilon [k]  = \mathbb{E}\left\{ {{{\left| {\hat s[k] - s[k]} \right|}^2}} \right\} \approx {\varepsilon ^{(1)}}[k] + {\varepsilon ^{(2)}}[k],
	\end{array}
\end{align}
where $\varepsilon _c^{(1)}[k] = {\left| {{e_c}[k]} \right|^2}{T_c}[k] - 2\Re \left\{ {{e_c}[k]{\bf{\hat h}}_{{\rm{equ}}}^H[k]{{\bf{f}}_{{\rm BB},c}}} \right\} + 1$ and ${\varepsilon ^{(1)}}[k] = {\big| {e[k]} \big|^2}T[k] - 2\Re \big\{ e[k] $ $ {\bf{\hat h}}_{{\rm{equ}}}^H[k]{{\bf{f}}_{{\rm{BB}}}}[k] \big\} + 1$ are the common and private MSEs calculated when the CSI errors are ignored, $\varepsilon _c^{(2)}[k] = {\left| {{e_c}[k]} \right|^2}\sum\limits_{m = 1}^K {{\bf{f}}_{{\rm{BB}}}^H[m]} {{\bf{R}}_e}{{\bf{f}}_{{\rm{BB}}}}[m]$ and ${\varepsilon ^{(2)}}[k] = {\left| {{e}[k]} \right|^2}\sum\limits_{m = 1,m \ne k}^K {{\bf{f}}_{{\rm{BB}}}^H[m]} {{\bf{R}}_e}{{\bf{f}}_{{\rm{BB}}}}[m]$ are the common and private MSEs caused by inter-stream interference introduced by the CSI errors,  and ${T_c}[k] = \sum\limits_{m = 0}^K \big| {\bf{\hat h}}_{{\rm{equ}}}^H[k]   {{\bf{f}}_{{\rm{BB}}}}[m] \big|^2  + \sigma _n^2 $ and $T[k] = \sum\limits_{m = 1}^K {{{\left| {{\bf{\hat h}}_{{\rm{equ}}}^H[k]{{\bf{f}}_{{\rm{BB}}}}[m]} \right|}^2} + \sigma _n^2} $ are  the average power of the received signal and the signal after removing the common data stream, respectively. Assuming that the CSI errors follow the i.i.d. complex Gaussian distribution, the autocorrelation matrix of the CSI errors can be expressed as ${{\bf{R}}_e} = \sigma _H^2{{\bf{I}}_K}$, where $\sigma _H^2$ denotes the MSE of the estimated equivalent CSI. By setting  $\frac{{\partial {\varepsilon _c}[k]}}{{\partial {e_c}[k]}} = 0$ and $\frac{{\partial {\varepsilon }[k]}}{{\partial {e}[k]}} = 0$, the minimum MSE (MMSE) equalizers are given as
\begin{align}
	\label{equ:e_MMSE}
\begin{array}{*{20}{l}}
	{e_c^{{\rm{MMSE}}}[k] = \dfrac{{{\bf{f}}_{{\rm{BB}},c}^H{{{\bf{\hat h}}}_{{\rm{equ}}}}[k]}}{{\left( {{T_c}[k] + \sigma _H^2\sum\limits_{m = 1}^K {{\bf{f}}_{{\rm{BB}}}^H[m]{{\bf{f}}_{{\rm{BB}}}}[m]} } \right)}},}\\
	{e_{}^{{\rm{MMSE}}}[k] = \dfrac{{{\bf{f}}_{{\rm{BB}}}^H[k]{{{\bf{\hat h}}}_{{\rm{equ}}}}[k]}}{{\left( {T[k] + \sigma _H^2\sum\limits_{m = 1,m \ne k}^K {{\bf{f}}_{{\rm{BB}}}^H[m]{{\bf{f}}_{{\rm{BB}}}}[m]} } \right)}}.}
\end{array}
\end{align}

Substituting (\ref{equ:e_MMSE}) into (\ref{equ:eps}), the MMSEs are given by
\begin{align}
	\label{equ:eps_MMSE}
\begin{array}{*{20}{l}}
	{\varepsilon _c^{{\rm{MMSE}}}[k] = 1 - \dfrac{{{{\left| {{\bf{f}}_{{\rm{BB}},c}^H{{{\bf{\hat h}}}_{{\rm{equ}}}}[k]} \right|}^2}}}{{\left( {{T_c}[k] + \sigma _H^2\sum\limits_{m = 1}^K {{\bf{f}}_{{\rm{BB}}}^H[m]{{\bf{f}}_{{\rm{BB}}}}[m]} } \right)}},}\\
	{\varepsilon _{}^{{\rm{MMSE}}}[k] = 1 - \dfrac{{{{\left| {{\bf{f}}_{{\rm{BB}}}^H[k]{{{\bf{\hat h}}}_{{\rm{equ}}}}[k]} \right|}^2}}}{{\left( {T[k] + \sigma _H^2\sum\limits_{m = 1,m \ne k}^K {{\bf{f}}_{{\rm{BB}}}^H[m]{{\bf{f}}_{{\rm{BB}}}}[m]} } \right)}}.}
\end{array}
\end{align}
Then, the SINRs of the common and private data streams can be rewritten as ${\gamma _c}[k] = 1/\varepsilon _c^{{\rm{MMSE}}}[k] - 1$ and $\gamma [k] = 1/\varepsilon _{}^{{\rm{MMSE}}}[k] - 1$, and the corresponding rates can be rewritten as $\widehat R_k^c =  - {\log _2}(\varepsilon _c^{{\rm{MMSE}}}[k])$ and $\widehat R_k^p =  - {\log _2}(\varepsilon _{}^{{\rm{MMSE}}}[k])$ \footnote{Since the BS cannot calculate the actual rates at the precoding stage and we have made some approximations in the calculation of the MSEs in formula (\ref{equ:eps}), we adopt $\widehat R_k^c$ and $\widehat R_k^p$ to represent the approximation of the rates, respectively.}. Since the logarithmic rate-MSEs relationship cannot be used directly for solving the rate optimization problem, we introduce the augmented weighted MSEs (WMSEs)
\begin{align}
	\label{equ:xi}
\begin{array}{l}
	{\xi _c}[k] = {\lambda _c}[k]{\varepsilon _c}[k] - {\log _2}\left( {{\lambda _c}[k]} \right),\\
	\xi [k] = \lambda [k]\varepsilon [k] - {\log _2}\left( {\lambda [k]} \right),
\end{array}
\end{align}
where ${\lambda _c}[k]$ and $\lambda [k]$ are the weights of the MSEs associated with the $k$-th UE. By setting $\frac{{\partial \xi _c[k]}}{{\partial {\lambda _c}[k]}} = 0$ and $\frac{{\partial \xi [k]}}{{\partial {\lambda}[k]}} = 0$, the optimal weights are given as
\begin{align}
	\label{equ:lambda_MMSE}
	{\lambda_c^{\rm MMSE}}[k] = {\left( {\varepsilon _c^{{\rm{MMSE}}}[k]} \right)^{ - 1}},\ \lambda^{\rm MMSE} [k] = {\left( {\varepsilon _{}^{{\rm{MMSE}}}[k]} \right)^{ - 1}}.
\end{align}

Substituting (\ref{equ:e_MMSE}) and (\ref{equ:lambda_MMSE}) into (\ref{equ:xi}), the rate-WMMSE relationships are established as 
\begin{align}
	\xi_c^{\rm MMSE}[k] = 1-\widehat R_k^c, \  \xi^{\rm MMSE}[k] = 1-\widehat R_k^p.
\end{align}

With the above rate-WMMSE relationships, the AWMMSE problem can be formulated as{\color{black}
\begin{align} \label{equ:WMMSE}
	\mathop{\rm minimize}\limits_{{\bf{F}}_{{\rm{BB}}},{\mathbf e},{\mathbf{\lambda}}}\quad &   {\xi^w} =    
	\left(   {\max} _k\left\{{\xi_c[k]}\right\}   +  {{{{\max} _k }}\left\{\xi[k]\right\}}   \right) \nonumber \\
	{\rm s.t.} \quad 
	& \|{\mathbf F}_{\rm RF}{\mathbf F}_{\rm BB}\|_F^2\le P_t,
\end{align}}
where ${\bf{e}} = {\left[ {{e_c}[1], \cdots ,{e_c}[K],e[1], \cdots ,e[K]} \right]^T}$ is the equalizer vector and ${\bf{\lambda }} = [ {\lambda _c}[1], \cdots ,{\lambda _c}[K],$ $\lambda [1], \cdots ,\lambda [K] ]^T$ is the weight vector. 
Although this AWMMSE problem is still non-convex, we can exploit the alternating optimization framework to decompose this non-convex problem into three convex subproblems by optimizing one of ${\bf{F}}_{{\rm{BB}}}$, ${\mathbf e}$, and ${\mathbf{\lambda}}$ separately while fixing the remaining two parts, as shown in Algorithm \ref{Alg1}.
Of these, ${\mathbf e}$ and ${\mathbf{\lambda}}$ can be solved in closed-forms using (\ref{equ:e_MMSE}) and (\ref{equ:lambda_MMSE}). As for ${\bf{F}}_{{\rm{BB}}}$, we can adopt CVX toolbox \cite{CVX} to solve for the optimal ${\bf{F}}_{{\rm{BB}}}$ while fixing ${\mathbf e}$ and ${\mathbf{\lambda}}$. The convergence proof for alternate optimization can be found in \cite{WMMSE_RSMA1}.

 \SetAlgoNoLine
\SetAlCapFnt{\normalsize}
\SetAlCapNameFnt{\normalsize}\
\begin{algorithm}[!t]

	\caption{AWMMSE-Based RSMA Digital Basedband Precoding}
	\begin{algorithmic}[1]
		\label{Alg1}
		\STATE \textbf{Initialize} the digital precoder ${\mathbf F}_{\rm BB}$ by using zero \\ forcing precoder;
		\FOR {$i=1$ to $I_1$}
		\STATE \textbf{Update} ${\mathbf e}$ and ${\mathbf \lambda}$ by using formula (\ref{equ:e_MMSE}) and (\ref{equ:lambda_MMSE}) \\ with fixed ${\mathbf F}_{\rm BB}$;
		\STATE \textbf{Update} ${\mathbf F}_{\rm BB}$ by solving problem (\ref{equ:WMMSE}) with fixed ${\mathbf e}$ and ${\mathbf \lambda}$;
		\ENDFOR

	\end{algorithmic}
\end{algorithm}
\vspace{-7mm}

\begin{figure*}[t]
	\vspace*{-7mm}
	\centering
	\includegraphics[width = 2 \columnwidth,keepaspectratio]
	{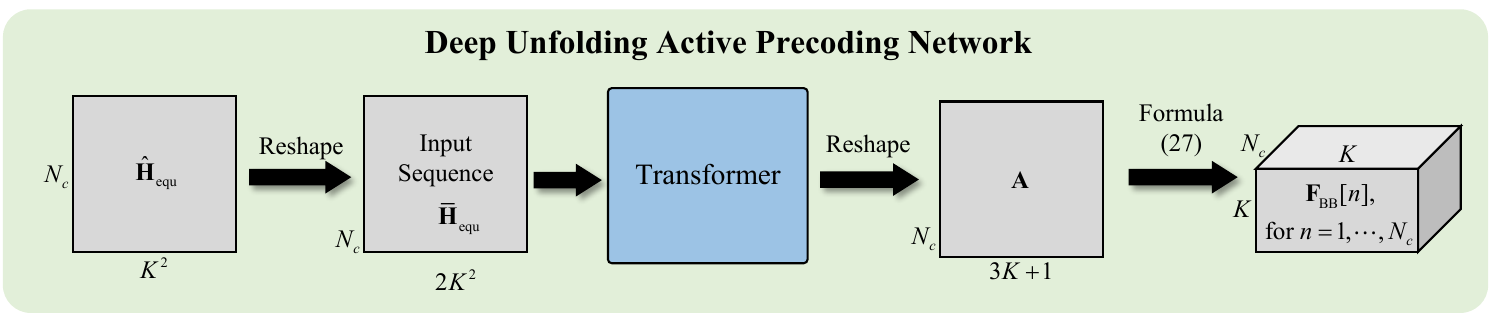}
	\captionsetup{font={footnotesize}, singlelinecheck = off, justification = raggedright,name={Fig.},labelsep=period}
	\vspace{-2mm}
	\caption{The proposed DFAPN for the design of the RSMA digital active precoding.}
	\label{fig:DFAPN}
	\vspace*{-4mm}
\end{figure*}

\subsection{Proposed DFAPN for the RSMA Digital Active Precoding at the BS}
The proposed AWMMSE algorithm can reduce the computational complexity compared to the method in \cite{WMMSE_RSMA1} by using approximation to avoid Monte-Carlo sampling. However, due to the $\max(\cdot)$ function in the optimization objective of problem (\ref{equ:WMMSE}), the proposed AWMMSE algorithm still cannot obtain a closed-form solution when updating ${\bf{F}}_{{\rm{BB}}}$ and needs to be computed iteratively by using the CVX tool, which increases the processing delay. To further reduce the computational complexity, we  relax problem (\ref{equ:WMMSE}) (i.e., replacing the original optimization objective $\left(   {\max} _k\left\{{\xi_c[k]}\right\}   +  {{{{\max} _k }}\left\{\xi[k]\right\}}   \right)$ with $\left( {\sum\nolimits_k {{\xi _c}[k]}  + \sum\nolimits_k {\xi [k]} } \right)$) and use Lemma 1 introduced in \cite{DL_unfloding1}, so as to obtain the update formulation of ${{{\bf{F}}_{{\rm{BB}}}}}$ via closed-form, which is given by
\begin{align} \label{equ:MDDL}
\left\{ {\begin{array}{*{20}{l}}
		\begin{array}{l}
			\;{\kern 1pt} {{\bf{f}}_{{\rm{BB}},c}} = {\left( {{b_c} + \sum _{m = 1}^K{c_c}[m]{{{\bf{\hat h}}}_{{\rm{equ}}}}[m]{\bf{\hat h}}_{{\rm{equ}}}^H[m]} \right)^{ - 1}}\\
			\quad\quad\quad\quad\sum\limits_{m = 1}^K {{{{\bf{\hat h}}}_{{\rm{equ}}}}[m]{a_c}[m],} 
		\end{array}\\
		\begin{array}{l}
			{{\bf{f}}_{{\rm{BB}}}}[k] = {\left( {b[k] + \sum _{m = 1}^Kc[m]{{{\bf{\hat h}}}_{{\rm{equ}}}}[m]{\bf{\hat h}}_{{\rm{equ}}}^H[m]} \right)^{ - 1}}\\
			\quad\quad\quad\quad{{{\bf{\hat h}}}_{{\rm{equ}}}}[k]a[k],\;{\kern 1pt} 1 \le k \le K,
		\end{array}
\end{array}} \right.
\end{align}
where {\color{black}
\begin{align}\label{equ:para}
	\left\{ {\begin{array}{*{20}{l}}
			{{a_c}[k] = {e_c}[k]{\lambda _c}[k],}\\
		\	{a[k] = e[k]\lambda [k],}\\
		\ \ \ \	{{b_c} = \frac{{\sigma _n^2}}{P_t}\sum\limits_{m = 1}^K {\left( {{\lambda _c}[k]{{\left| {{e_c}[k]} \right|}^2} + \lambda [k]{{\left| {e[k]} \right|}^2}} \right),} }\\
		\ \ 	{b[k] = {b_c} + \sigma _H^2\sum\limits_{m = 1,m \ne k}^K {\left( {{\lambda _c}[k]{{\left| {{e_c}[k]} \right|}^2}} \right).} }
	\end{array}} \right.
\end{align}}
Consider  index $n$ of the subcarrier, then the values of the digital precoder $\mathbf{F}_{\rm BB}[n]$ depends on ${\bf{A}} = \{ {a_c}[k,n],a[k,n],{b_c}[n],b[k,n],\ {\rm for}\ k = 1, \cdots ,K,\ n = 1, \cdots,{N_c} \}\in\mathbb{C}^{N_c\times (3K+1)}$, which can be calculated by formula (\ref{equ:e_MMSE}), (\ref{equ:lambda_MMSE}), and (\ref{equ:para}).

The AWMMSE algorithm derived in the previous subsection is based on the assumption that the CSI errors are complex Gaussian distributed and independent of the CSI, which may not exactly match the actual scenario. 
Besides, the objective function in (\ref{equ:WMMSE}) is relaxed for the derivation of the closed-form update formulation (\ref{equ:para}), which may further degrade the performance.
To this end, we propose a model-driven DFAPN by deep unfolding the proposed AWMMSE scheme for the RSMA digital active precoding, as shown in Fig. \ref{fig:DFAPN}. Specifically, the proposed DFAPN takes the estimated equivalent channel ${{\bf{\widehat H}}_{{\rm{equ}}}} \in \mathbb{C} {^{{N_c}\times{K^2}  }}$ as the input, i.e.,
\begin{align}
	{{\bf{\widehat H}}_{{\rm{equ}}}} = &\big[ {\rm{vec}}\left( {\left[ {{{{\bf{\hat h}}}_{{\rm{equ}}}}[1,1], \cdots ,{{{\bf{\hat h}}}_{{\rm{equ}}}}[K,1]} \right]} \right), \cdots ,\nonumber \\ & {\rm{vec}}\left( {\left[ {{{{\bf{\hat h}}}_{{\rm{equ}}}}[1,{N_c}], \cdots ,{{{\bf{\hat h}}}_{{\rm{equ}}}}[K,{N_c}]} \right]} \right) \big]^H.
\end{align}
 Then, the DFAPN converts ${{\bf{\widehat H}}_{{\rm{equ}}}}$ into a 1D real-valued input sequence ${{\bf{\bar H}}_{{\rm{equ}}}}\in\mathbb{C} {^{{N_c} \times {2K^2}  }}$, i.e.,
\begin{align}
	\left\{ {\begin{array}{*{20}{l}}
		\ \ \ \ \ \ 	{{{\left[ {{{{\bf{\bar H}}}_{{\rm{equ}}}}} \right]}_{[:,1:{K^2}]}} = \Re \left\{ {{{{\bf{\widehat H}}}_{{\rm{equ}}}}} \right\},}\\
			{{{\left[ {{{{\bf{\bar H}}}_{{\rm{equ}}}}} \right]}_{[:,1 + {K^2}:2{K^2}]}} = \Im \left\{ {{{{\bf{\widehat H}}}_{{\rm{equ}}}}} \right\},}
	\end{array}} \right.
\end{align}
which is then processed by a Transformer to output the key parameters $\bf{A}$ in the proposed AWMMSE scheme. Formula (\ref{equ:MDDL}) is then used to obtain the ${\rm F}_{\rm BB}[n]$, for $n=1,\cdots,N_c$. Finally, we impose the power constraint on the digital precoder, i.e.,{\color{black}
\begin{align}
	{{\bf{F}}_{{\rm{BB}}}}[n] = \min &\left( {\sqrt P_t ,{{\left\| {{\mathbf F}_{\rm RF}{{\bf{F}}_{{\rm{BB}}}}[n]} \right\|}_F}} \right)\frac{{{{\bf{F}}_{{\rm{BB}}}}[n]}}{{{{\left\| {{\mathbf F}_{\rm RF}{{\bf{F}}_{{\rm{BB}}}}[n]} \right\|}_F}}},\; \nonumber \\ &1 \le n \le {N_c}.
\end{align}}

The above RSMA digital active precoding process can be written as
\begin{align}
	\left\{ {{{\bf{F}}_{{\rm{BB}}}}[1], \cdots ,{{\bf{F}}_{{\rm{BB}}}}[{N_c}]} \right\} = {\cal P}({\widehat {\bf{H}}_{{\rm{equ}}}};{{\bf{W}}_{{\rm{RP}}}}),
\end{align}
where ${\cal P}(\cdot;{\mathcal W}_{\rm RP})$ is the mapping function from ${{\bf{\widehat H}}_{{\rm{equ}}}}$ to the digital precoder, and ${\mathcal W}_{\rm RP}$ is the learnable neural network parameters.

In summary, the learnable parameters of the proposed DL-based precoding scheme include the learnable parameter set ${\mathcal W}_{\rm RR}$ of the RRN and the learnable parameter set ${\mathcal W}_{\rm RP}$ of the DFAPN. 
We choose the negative ARWU as the loss function, i.e.,
\begin{align} \label{equ:loss2}
	{L_{{\rm{E2E}}}^{(1)}} &= -\sum\limits_{n = 1}^{{N_c}}{R^w_n} \nonumber \\ &=    
	-\sum\limits_{n = 1}^{{N_c}}\left(   {{{{\min}_k }}\left\{ {R_{k,n}^p} \right\}}  +  {{{{\min}_k }}\left\{R_{k,n}^c\right\}}   \right),
\end{align}
thus performing E2E DL training on the proposed RRN and DFAPN to achieve better performance.

\section{Proposed DL-Based CSI Acquisition Scheme} \label{S:CSI_acu}
\begin{figure*}[t]
	\vspace*{-5mm}
	\centering
	\includegraphics[width = 2 \columnwidth,keepaspectratio]
	{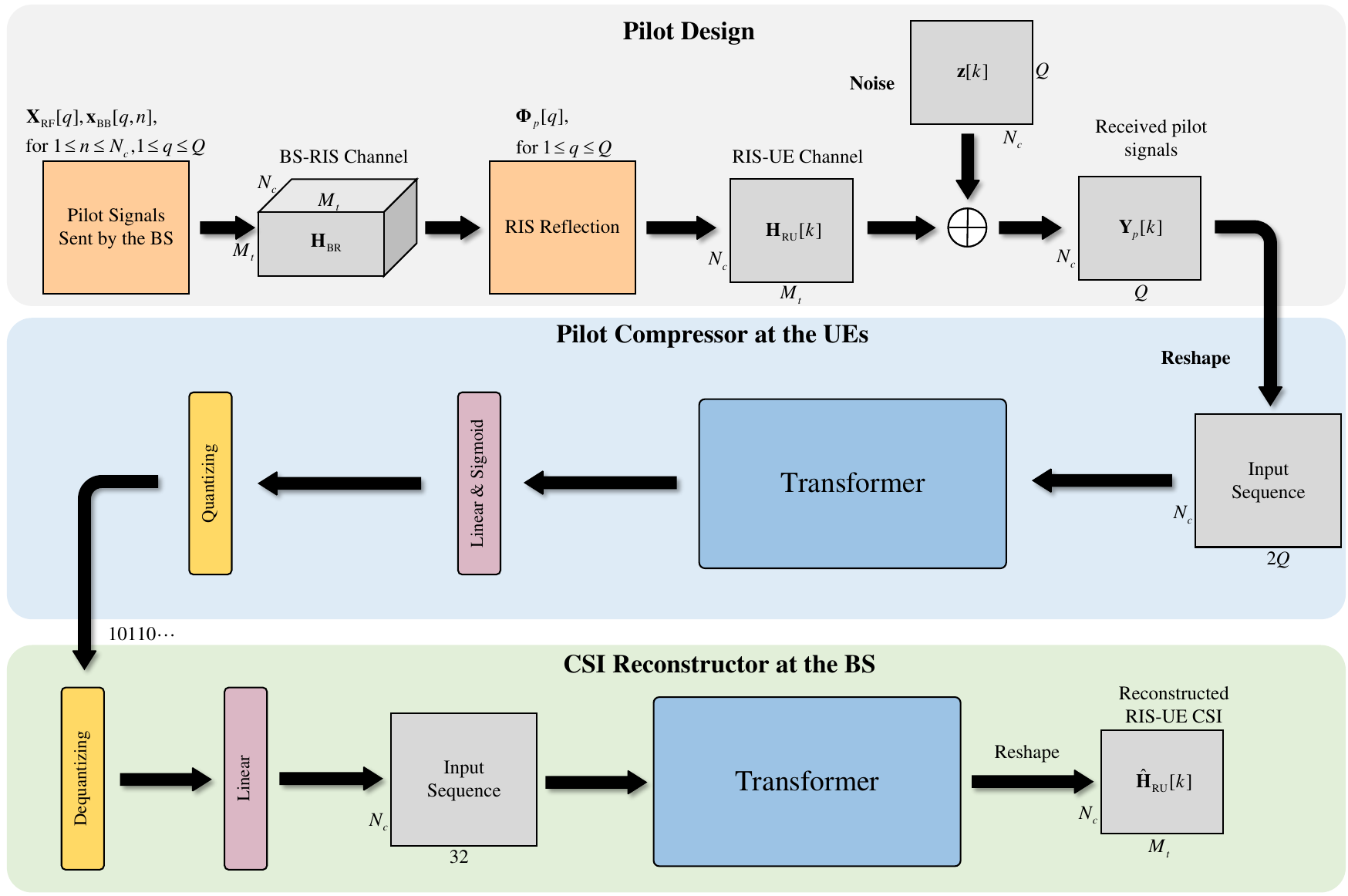}
	\captionsetup{font={footnotesize}, singlelinecheck = off, justification = raggedright,name={Fig.},labelsep=period}
	\caption{The proposed CAN for the joint optimization of the downlink pilot signals at the BS and the RIS, uplink CSI feedback at the UEs, and the channel reconstruction at the BS.}
	\label{fig:CAN}
	\vspace*{-5mm}
\end{figure*}

In this section, we propose an E2E neural network, denoted by CAN, to jointly design the downlink pilot signals, uplink CSI feedback at the UEs, and channel reconstruction at the BS. As shown in Fig. \ref{fig:CAN}, the proposed CAN consists of the downlink pilot signals, a  pilot compressor at the UEs, and a CSI reconstructor at the BS. 
In following, we firstly present the processing procedure of the proposed scheme and then describe how to model such a CSI acquisition procedure as an E2E neural network.

\subsection{Processing Procedure and Problem Formulation for the CSI Acquisition}
To estimate the RIS-UE channels, the BS needs to send multiple downlink pilot OFDM symbols to the UEs through the reflection caused by the RIS, and reconstruct the downlink RIS-UE CSI according to the feedback information from the UEs. 
At the downlink pilot transmission stage, we assume that the BS activates all the RF chains to transmits $Q$ consecutive pilot OFDM symbols.
For the $k$-th UE, the downlink pilot signals received at the $n$-th subcarrier of the $q$-th pilot OFDM symbol, denoted by $y_p[q,k,n]\in\mathbb{C}$, can be expressed as
\begin{align} 
	y_p[q,k,n]= & {\mathbf h}_{\rm RU}^H[k,n]{\mathbf \Phi}_{p}[q]{\mathbf H}_{\rm BR}^H[n]{\mathbf X}_{\rm RF}[q]{\mathbf x}_{\rm BB}[q,n] \nonumber \\ & + z[q,k,n],
\end{align}
where ${z}[{q,k,n}]\sim {\cal CN}\left( {0},\sigma_n^2 \right)$  is the AWGN, ${\mathbf \Phi}_{p}[q]\in\mathbb{C}^{M_r\times M_r}$ is the reflecting matrix of the RIS, ${\mathbf x}_{\rm BB}[q,n]\in \mathbb{C}^{K\times 1}$ and ${\mathbf X}_{\rm RF}[q]\in \mathbb{C}^{M_b\times K}$ are the digital baseband  pilot signals and analog RF  pilot signals sent by the BS for the $q$-th pilot OFDM symbol, respectively.
Moreover, ${\mathbf \Phi}_{p}[q]$, ${\mathbf X}_{\rm RF}[q]$, and ${\mathbf x}_{\rm BB}[q,n]$ should satisfy 
the unit modulus constraints and power constraint, i.e. ${\mathbf \Phi}_{p}[q]\in {\cal{F_{\rm RIS}}}$, ${\mathbf X}_{\rm RF}[q]\in {\cal{F_{\rm RF}}}$, and {\color{black}$|{\mathbf X}_{\rm RF}[q]{\mathbf x}_{\rm BB}[q,n]|_2^2\le P_t$}.
By combining the received pilot signals in $Q$ pilot OFDM symbols at the $k$-th UE at the $n$-th subcarrier, the received pilot signals can be aggregately expressed as
\begin{equation} \label{equ:pilot}
	{\bf{y}}_p[k,n] = {\bf{X}}[n]{\bf{h}}_{\rm RU}[k,n] + {\bf{z}}[k,n],
\end{equation}
where ${\bf{X}}[n] = [\bf{\Phi _p}[1]{\bf{H}}_{{\rm{BR}}}^H[n]{{\bf{X}}_{{\rm{RF}}}}[1]{{\bf{x}}_{{\rm{BB}}}}[1,n], \cdots ,\bf{\Phi }_p[Q]$ ${\bf{H}}_{{\rm{BR}}}^H[n]{{\bf{X}}_{{\rm{RF}}}}[Q]{{\bf{x}}_{{\rm{BB}}}}[Q,n]]^H\in\mathbb{C}^{Q\times M_r}$, ${\bf{y}}_p[k,n] = \left[ y_p[1,k,n],\cdots,y_p[Q,k,n] \right]^H\in\mathbb{C}^{Q\times 1}$, and ${\bf{z}}[k,n]=\left[ z[1,k,n],\cdots,z[Q,k,n] \right]^H\in\mathbb{C}^{Q\times 1}$. 
By combining all the subcarriers, we denote the total pilot signals received at the $k$-th UE as ${\mathbf Y}_p[k]=\big[ {\bf{y}}_p[k,1],\cdots$ ${,\bf{y}}_p[k,N_c] \big]^H\in\mathbb{C}^{N_c\times Q}$.
At the uplink CSI feedback stage, the $k$-th UE needs to extract the CSI from the received pilot signals ${\mathbf Y}_p[k]$, quantize them into $B$ bits, and feed them back to the BS. We can mathematically express this process as
\begin{align}
	{\mathbf q}[k] ={\cal Q}({\mathbf Y}_p[k])\in \mathbb{R}^{B\times 1},
\end{align}
where ${\cal Q}(\cdot)$ compresses and quantizes the received pilot signals ${\mathbf Y}[k]$ into a feedback bit vector ${\mathbf q}[k]$.

Based on the feedback bit vector ${\mathbf q}[k]$ from the $k$-th UE, the BS needs to reconstruct the downlink RIS-UE CSI. This process can be expressed as
\begin{align}
	\widehat{\mathbf H}_{\rm RU}[k] ={\cal C}({\mathbf q}[k]),
\end{align}
where ${\cal C}(\cdot)$ represents the mapping function from the feedback bit vector ${\mathbf q}[k]$ to the reconstructed downlink RIS-UE CSI $\widehat{\mathbf H}_{\rm RU}[k]$.

Based on the above processing procedure, we consider to minimize the normalized mean square error (NMSE) of the reconstructed CSI. The joint design of the downlink pilot training, uplink CSI feedback, and channel reconstruction can be formulated as{\color{black}
\begin{align}
	\label{equ:P1}
	\mathop{\rm minimize}\limits_{{\mathcal W},{\cal Q(\cdot)},{\cal C(\cdot)}}\quad & {\rm NMSE} = \dfrac{1}{K} \sum\limits_{k=1}^K 
	\dfrac{\left\|\widehat{\mathbf H}_{\rm RU}[k]-{\mathbf H}_{\rm RU}[k] \right\|_F^{2}}{\left\|{\mathbf H}_{\rm RU}[k] \right\|_F^{2} }  \nonumber \\
	{\rm s.t.} \quad\quad &  {\mathbf q}[k] ={\cal Q}({\mathbf Y}_p[k]),\forall k,  \nonumber \\
	& \widehat{\mathbf H}_{\rm RU}[k] ={\cal C}({\mathbf q}[k]),\forall k,n,  \nonumber \\
	& {\mathbf \Phi}_{p}[q]\in {\cal{F_{\rm RIS}}},\forall q,  \nonumber \\
	& {\mathbf X}_{\rm RF}[q]\in {\cal{F_{\rm RF}}},\forall q, \nonumber \\
	& |{\mathbf X}_{\rm RF}[q]{\mathbf x}_{\rm BB}[q,n]|_2^2\le P_t, \forall q,n,
\end{align}}
where $ {\mathcal W}=\left\{{{\mathbf \Phi}_{p}[q],{\mathbf X}_{\rm RF}[q],{\mathbf x}_{\rm BB}[q,n],\forall q,n}\right\}$ is the set of the pilot signals sent by the BS and the reflecting matrix at the RIS that need to be designed at the pilot transmission stage. 

Jointly optimizing the above pilot design, uplink CSI feedback at the UEs, and CSI reconstruction at the BS is challenging since this optimization problem faces complicated constraints and a large number of variables. Conventional schemes usually consider optimizing these modules independently, which faces excessive pilot and feedback signaling overhead. To this end, by modeling the pilot design,  uplink CSI feedback at the UEs, and CSI reconstruction at the BS  as an E2E neural network, we propose the data-driven DL-based CAN to achieve the joint optimization.

\subsection{Downlink Pilot Design}
At the downlink pilot transmission stage represented in (\ref{equ:pilot}), the parameters that need to be designed include digital baseband pilot signals ${{\bf{x}}_{{\rm{BB}}}}[q,n]$, analog RF pilot signals ${{\bf{X}}_{{\rm{RF}}}}[q]$, and RIS reflecting matrix ${\mathbf \Phi}_{p}[q]$, for $1\leq n \leq N_c$, $1\leq q \leq Q$. 
Different from the existing DL schemes \cite{Ma_tvt,Ma_JSAC} that directly model the pilot signals as a fully-connected layer, we consider a more complicated RIS-aided MIMO-OFDM system, where the digital baseband pilot signals are frequency-selective, while the analog RF pilot signals and RIS reflecting matrix are frequency-flat that satisfy the unit modulus constraints.
Therefore, we take the digital baseband pilot signals ${{\bf{x}}_{{\rm{BB}}}}[q,n]$, the phase values ${\mathbf \Theta}_{\rm RF}^p[q]\in\mathbb{R}^{M_b\times K}$ of the analog RF pilot signals, and the phase values ${\mathbf \Theta}_{\rm RIS}^p[q]\in\mathbb{R}^{M_r\times M_r}$ of the RIS reflecting matrix as trainable parameters in the CAN, which can be learned and determined at the DL training stage.


Then, we directly apply the complex exponent function to the phase matrices ${\mathbf \Theta}_{\rm RF}^p$ and ${\mathbf \Theta}_{\rm RIS}^p$ to generate
\begin{align}
	\left\{ {\begin{array}{*{20}{l}}
		\ \ 	{{\bf{X}}_{{\rm{RF}}}}[q] = \exp (1{\rm{j}} \cdot {\bf{\Theta }}_{{\rm{RF}}}^p[q])/\sqrt{M_b},\forall q,\\
			{{{\left[ {{\bf{\Phi }}_{p}[q]} \right]}_{i,j}} = \exp (1{\rm{j}} \cdot {{\left[ {{\bf{\Theta }}_{{\rm{RIS}}}^p[q]} \right]}_{l,l}}),\ \ {\rm{  if \ }} i = j,}\\
			{{{\left[ {{\bf{\Phi }}_{p}[q]} \right]}_{i,j}} = 0,\ \ {\rm{  others}}},
	\end{array}} \right.
\end{align}
such that the unit modulus constraints of the analog pilot signals ${\mathbf X}_{\rm RF}[q]\in {\cal{F_{\rm RF}}}$ and the RIS reflecting matrix ${\mathbf \Phi}_{p}[q]\in {\cal{F_{\rm RIS}}}$ can be satisfied.


Finally, we consider the power normalization of the digital pilot signals to satisfy the power constraint, i.e., {\color{black}
\begin{align}
	{\mathbf x}_{\rm BB}[q,n] = \dfrac{\sqrt{P_t}{\mathbf x}_{\rm BB}[q,n]}{|{\mathbf X}_{\rm RF}[q]{\mathbf x}_{\rm BB}[q,n]|_2},\forall q,n.
\end{align}}

\subsection{Uplink CSI Feedback}
At the uplink CSI feedback stage, we model the process that the $k$-th UE extracts the CSI from the received pilot signals and feeds it back to the BS as a Transformer-based pilot compressor at the UEs, as shown in Fig. \ref{fig:CAN}. 
 To handle the input of the proposed compressor, i.e., a 2D complex-valued matrix ${\mathbf Y}_p[k]$, we reshape the received pilot signal into a 1D real-valued input sequence  $\bar{\mathbf Y}_{p}[k]\in\mathbb{R}^{N_c\times 2Q}$, i.e.,
\begin{equation}
	\begin{cases}
	\ \ \ \ 	\left[ \bar{\mathbf Y}_{p}[k]\right]_{[:,1:Q]}=\Re\{{{\mathbf Y}_{p}[k]}\},\\
		\left[ \bar{\mathbf Y}_{p}[k]\right]_{[:,1+Q:2Q]}=\Im\{{{\mathbf Y}_{p}[k]}\},
	\end{cases}
\end{equation}
which is then processed by a Transformer. Afterwards, the sequence is compressed into a codeword by a fully-connected linear layer and a sigmoid activation function. Finally, the compressed codeword is quantized into $B$ feedback bits by a quantization layer.

Based on the above process, the feedback bit vector can be expressed as
\begin{align}
	{\mathbf q}[k] ={\cal Q}({\mathbf Y}_{\rm p}[k];{\mathcal W}_{\rm PC}),
\end{align}
where ${\cal Q}(\cdot;{\mathcal W}_{\rm PC})$ is the mapping function from the received pilot signals ${\mathbf Y}_{\rm p}[k]$ to the feedback vector ${\mathbf q}[k]$ and ${\mathcal W}_{\rm PC}$ is the learnable neural network parameters.

\subsection{Channel Reconstruction}
At the BS, the downlink RIS-UE CSI can be reconstructed based on the received feedback bit vector ${\mathbf q}[k]$, and we assume that there is no feedback error from the UEs to the BS. We model the channel reconstruction process as a Transformer-based CSI reconstructor at the BS as shown in the lower-half Fig. \ref{fig:CAN}.
The received feedback bit vector is first fed into a dequantization layer, and then be converted into the input sequence of the Transformer through a fully-connected linear layer. 
Finally, the Transformer further extracts the features of the input sequence and outputs the downlink reconstructed RIS-UE CSI $\widehat{\mathbf H}_{\rm RU}[k]$. The above channel reconstruction process can be written as
\begin{align}
	\widehat{\mathbf H}_{\rm RU}[k] ={\cal R}({\mathbf q}[k];{\mathcal W}_{\rm CR}),
\end{align}
where ${\cal Q}(\cdot;{\mathcal W}_{\rm CR})$ is the mapping function from ${\mathbf q}[k]$ to the reconstructed CSI  $\widehat{\mathbf H}_{\rm RU}[k]$, and ${\mathcal W}_{\rm CR}$ is the learnable neural network parameters.

In summary, the learnable parameters of the proposed CAN include the pilot signals (i.e., ${\mathbf \Theta}_{\rm RF}^p[q]$, ${\mathbf \Theta}_{\rm RIS}^p[q]$, and ${\mathbf x}_{\rm BB}[q,n]$, $\forall q,n$), the learnable parameter set ${\mathcal W}_{\rm PC}$ of the pilot compressor at the UEs, and the learnable parameter set ${\mathcal W}_{\rm CR}$ of the CSI reconstructor at the BS. By using NMSE as the loss function, i.e.,
\begin{align}
	L_{{\rm{E2E}}}^{(2)} = {\rm NMSE} = \dfrac{1}{K} \sum\limits_{k=1}^K 
	\dfrac{\left\|\widehat{\mathbf H}_{\rm RU}[k]-{\mathbf H}_{\rm RU}[k] \right\|_F^{2}}{\left\|{\mathbf H}_{\rm RU}[k] \right\|_F^{2} } ,
\end{align}
we can perform E2E DL training on the CAN to obtain the above learnable parameters.

\section{Numerical Results}\label{S:num}
\subsection{Simulation Schemes}

For the evaluation of the CSI acquisition performance, we simulate and compare the following schemes.

\begin{itemize}
	\item {{\em Perfect estimation \& {\textbf{csiNet}}}: Consider the perfect CSI at the UEs, CSI estimation is not required and CSI feedback is performed using a CNN-based CSI feedback scheme, i.e., csiNet \cite{csinet1}. }
\end{itemize}

\begin{itemize}
	\item {{\em{\textbf{GMMV-SOMP/BSOMP/AMP/LAMP}} \& Perfect feedback}: In this case, we consider the UEs utilize the generalized MMV (GMMV)-simultaneous OMP (SOMP) \cite{GMMV_SOMP}, GMMV-blocked SOMP (BSOMP) \cite{GMMV_BSOMP}, GMMV-AMP \cite{GMMV_AMP}, or GMMV-LAMP\footnote{Since the MMV scenario (i.e., the transmit pilot signals are the same at different subcarriers) is considered in \cite{Ma_JSAC} while the GMMV scenario (i.e., the transmit pilot signals are different at different subcarriers) is considered in this paper, we consider extending the MMV-LAMP network proposed in \cite{Ma_JSAC} to a GMMV-LAMP network by replacing the frequency-flat measurement matrix and the learnable parameters with frequency-selective parameters.} \cite{Ma_JSAC} for downlink RIS-UE CSI estimation, which is then assumed to be perfectly fed back to the BS.}
\end{itemize}

\begin{itemize}
	\item {{\em{\textbf{GMMV-SOMP/BSOMP/AMP/LAMP}} \& {\textbf{csiNet}}}:
		{In this case, we consider the GMMV-SOMP, GMMV-BSOMP, GMMV-AMP, or GMMV-LAMP is used to estimate the downlink RIS-UE CSI, which is then fed back to the BS via csiNet \cite{csinet1}, i.e., including the CSI compressor at the UEs and CSI reconstructor at the BS. }}
\end{itemize}

\begin{itemize}
	\item {{\em{Proposed {\textbf{CAN}}}}:
		{In this case, the donwlink pilot signals generated by the BS and RIS, the pilot compressor at the UEs, and the CSI reconstructor at the BS are joint optimized in an E2E training manner. }}
\end{itemize}

Next, we present the simulation schemes for assessing the performance of precoding.
For a fair comparison, all the precoding schemes presented below are based on the CSI obtained by the proposed CAN. Besides, all the precoding schemes presented below adopt the MF-based analog active precoding introduced in Section~\ref{s:MF}.

\begin{itemize}
	\item {{\em {Beam alignment \& {\textbf {AWMMSE/power allocation } (RSMA)}}}:
		{In this case, we consider to use the beam alignment-based RIS reflecting matrix design, i.e., each subarray of the RIS selects a user and performs beam alignment according to the RIS-UE CSI at the central subcarrier\footnote{The design of the RIS reflecting matrix based on beam alignment can be expressed as ${\left[ {{\rm{diag}}\left( {\bf{\Phi }} \right)} \right]_{[(k - 1){M_{{\rm{sub}}}^r} + 1:k{M_{{\rm{sub}}}^r}]}} = \big[{{\bf{\hat h}}_{{\rm{RU}}}}[k,{N_c}/2]/| {{{{\bf{\hat h}}}_{{\rm{RU}}}}[k,{N_c}/2]} |\big]_{[(k - 1){M_{{\rm{sub}}}^r} + 1:k{M_{{\rm{sub}}}^r}]}$, where each subarray of the RIS is considered to select one UE to concentrate and reflect the signal energy to it, while ignoring interference to other UEs.}. Besides, we adopt the AWMMSE scheme derived in Section \ref{S:precoding} or the power allocation scheme\footnote{The common and private parts of the RSMA digital baseband precoder in the power allocation scheme are the matched beamforming (MBF) precoder and the RZF precoder, respectively, and the power control factor $\alpha$ is used to control the power ratio of the common and private parts.  The power allocation factor $\alpha$ in \cite{RSMA_Power3} is derived under the assumption that the CSI errors are independent of the CSI and follow i.i.d. complex Gaussian distribution, which does not match the actual CSI errors introduced by the proposed CAN. As such, we exhaust $\alpha$ at small intervals and calculate the ARWU based on the perfect CSI to find the power allocation factor that maximizes the ARWU. Although this baseline scheme can find the near-optimal $\alpha$ by exhaustive enumeration and does not require complicated mathematical derivation, this scheme cannot be applied in practice since we cannot obtain the perfect CSI  in practice to calculate the actual achievable rate to select the optimal power allocation factor. Therefore, this scheme is only used as an idealistic baseline scheme.} in \cite{RSMA_Power3} for the RSMA digital active precoding. }}
	
	\item {{\em {Beam alignment \& {\textbf {RZF }}(SDMA)}}:
		{In this case, we consider to use beam alignment-based RIS reflecting matrix design and regularized zero forcing (RZF) precoder for the SDMA digital active precoding. }}
	
	\item {{\em {Beam alignment \& Proposed{\textbf { DFAPN }}(RSMA/SDMA)}}:
		{In this case, we consider to use the beam alignment-based RIS reflecting matrix design and the proposed DFAPN for the RSMA or SDMA digital active precoding. As for the SDMA scenario, we only need to fix the power allocated to the common data stream to 0.}}
	
	\item {{\em {Proposed \textbf{RRN} \& Proposed{\textbf { DFAPN }}(RSMA/SDMA)}}:
		{In this case, we consider to use the proposed RRN for RIS reflecting matrix design and the proposed DFAPN for the RSMA or SDMA digital active precoding. The proposed RRN and DFAPN are jointly trained with the loss function (\ref{equ:loss2}).}}
	
\end{itemize}

\subsection{Channel Samples for Network Training}

\begin{figure*}[t]
	\vspace*{-8mm}
	\centering
	\includegraphics[width = 1.8 \columnwidth,keepaspectratio]
	{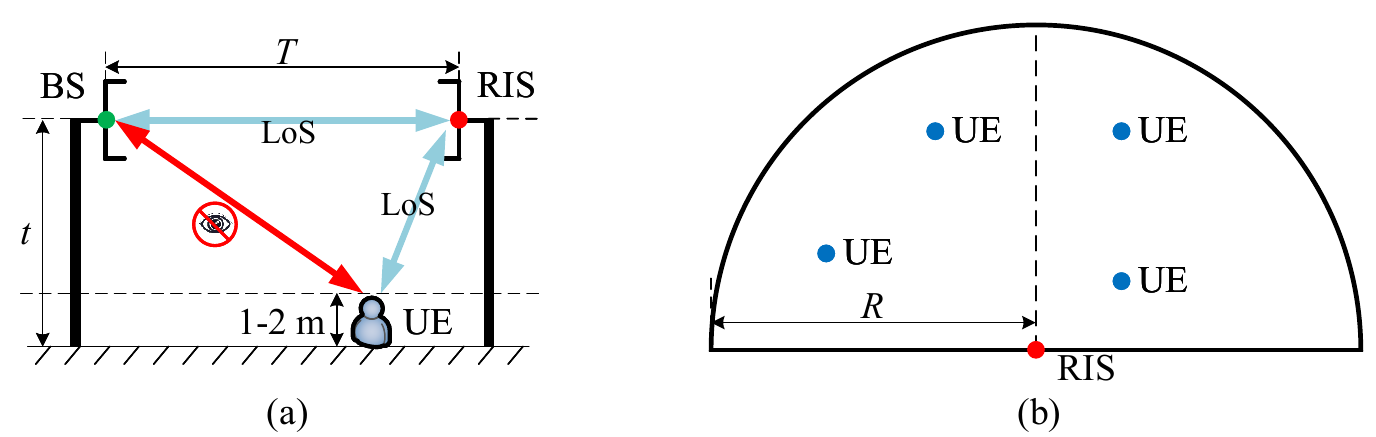}
	\captionsetup{font={footnotesize}, singlelinecheck = off, justification = raggedright,name={Fig.},labelsep=period}
	\vspace{-2mm}
	\caption{RIS-aided Tera-Hertz massive MIMO-OFDM systems. (a) Side view; (b) Top view.}
	\label{fig:simu_SYS}
	\vspace*{-4mm}
\end{figure*}

{\color{black}We consider that the carrier frequency is $f_c=150$ GHz (i.e., the carrier wavelength is $\lambda_c = 2$ mm), the bandwidth is $B_W=384$ MHz, the number of subcarriers is $N_c=64$, and the noise power spectral density is $-174$ dBm/Hz.
Both the BS and the RIS are equipped with a planar array on the $yz$-plane, and both have $K_y=2$ and $K_z=2$ subarrays uniformly distributed with spacing of $D_y=100\ \lambda_c=0.1$ m and $D_z=0.1$ m on the $y$ and $z$ axes, respectively.  Each subarray of the BS and RIS has $M_y=8$ and $M_z=8$ array elements uniformly distributed at half carrier wavelength $d=\lambda_c/2=1$ mm on the y and z axes, respectively.
Thus $K=K_yK_z=4$ is the total number of subarrays and $M_b=M_r=KM_yM_z=256$ is the total number of array elements at the BS and the RIS.
Furthermore, we consider the system model as shown in Fig. \ref{fig:simu_SYS}, where the deployment heights of the BS and the RIS are both $t=10$ m, the height of each UE is randomly distributed between 1-2 m, the normal directions of the planar arrays of the BS and the RIS point towards each other with a distance of $T=20$ m, and the RIS simultaneously serve $K$ active UEs distributed within a sector of radius $R=10$ m and central angle $180^{\circ}$.}\footnote{In this paper, we set $M_b$ and $M_r$ as the same value of 256 to simplify the calculation. In fact, it is only necessary to satisfy ${D_y} = \sqrt {{{{\lambda _c}T}}/{{{K_y}}}}$ and ${D_z} = \sqrt {{{{\lambda _c}T}}/{{{K_z}}}}$ to ensure single LoS-path multi-stream multiplexing in the LoS-MIMO architecture \cite{LoS_MIMO}, while parameters including $M_b$, $M_r$, $K$, etc. can be adjusted under the above constraint.}

Based on the above parameters, we can generate the corresponding channel samples according to (\ref{equ:h_BR}) and (\ref{equ:h_RU}). 
Note that we consider the static channel between the BS and the RIS, thus we only need to use (\ref{equ:h_BR})  to generate ${\mathbf H}_{\rm BR}[n]$ once. 
However, for the channel between the RIS and the UEs, since the UEs' locations are not fixed, we need to randomly sample the UEs' locations multiple times, thereby generating a large number of channel samples between the RIS and the UEs to train the proposed CAN, RRN, and DFAPN.



\subsection{Training Settings}
We set the number of layers to $U=6$  and the dimension of linear embedding to $d_{\rm model}=256$ in each Transformer.
Besides, we exploit the open-source DL framework PyTorch to train and validate the proposed neural network on a computer with dual Nvidia GeForce GTX 2080Ti GPUs. 
At the training stage, we adopt the Adam optimizer \cite{Adam} to update the network parameters and set the batch size of the training set to 128.
Besides, we vary the learning rate at each training step to accelerate the convergence of the proposed networks by using  warm-up strategy introduced in \cite{Transformer}, according to the formula:
\begin{align} \label{equ:warm}
	{{\rm{LR}}[p]} = d_{{\rm{model}}}^{ - 0.5} \cdot \min \left( p^{ - 0.5},p \cdot P^{ - 1.5} \right),
\end{align}
where ${{\rm{LR}}[p]}$ is the learning rate at the $p$-th training step and $P$ is the number of training steps for warm-up. We use $P=4000$. During the training, we also use the early-stop strategy in \cite{csinet1} to monitor the generalization performance of the proposed networks on the validation set, i.e., we can stop the training and retain the network parameters with the best generalization performance when the generalization performance has not increased for a relatively long time.

\subsection{Performance Comparison of Different CSI Acquisition Schemes}

\begin{figure*}[t]
	\vspace{-6mm}
	\captionsetup{font={footnotesize}, singlelinecheck = off, justification = raggedright,name={Fig.},labelsep=period}
	\centering
	\centering
	\hspace{-2mm}
	\begin{minipage}[t]{0.49\linewidth}
		\centering
		\includegraphics[scale=0.49]{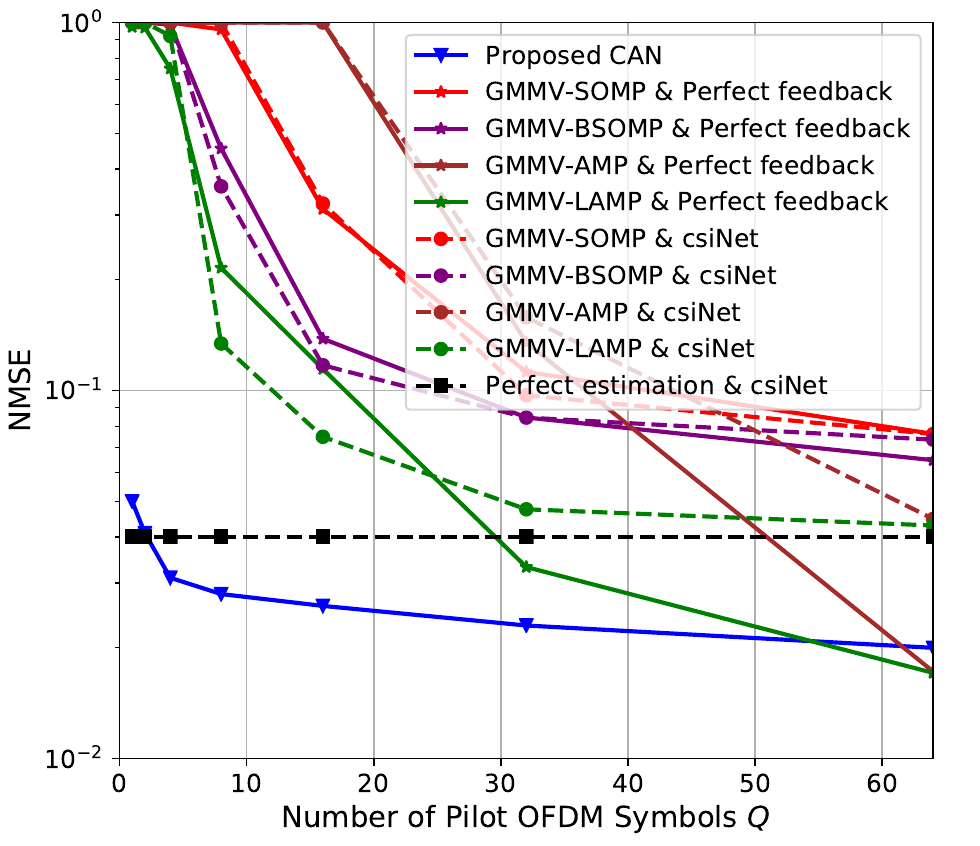}
		\vspace{-1.0mm}
		\captionsetup{font={footnotesize, color = {black}}, singlelinecheck = off, justification = raggedright,name={Fig.},labelsep=period}
		\caption{NMSE performance of different schemes versus the number of pilot OFDM symbols $Q$  ($B=32$, $P_t=40 \ {\rm dBm}$).}
		\label{fig:7}
	\end{minipage}
	\hfill
	\begin{minipage}[t]{0.49\linewidth}
		\centering
		\includegraphics[scale=0.49]{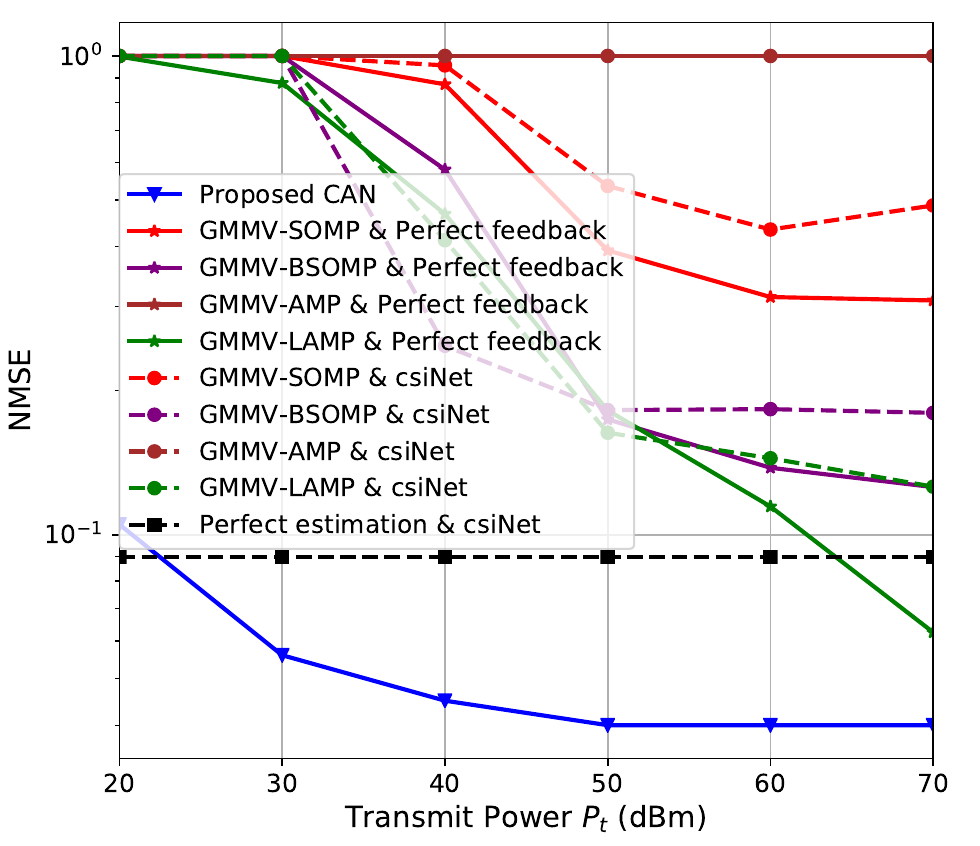}
		\vspace{-1.0mm}
		\caption{NMSE performance of different schemes versus the transmit power $P_t$  ($B=32$, $Q=16$).}
		\label{fig:8}
	\end{minipage}
	
	\vspace{-3.0mm}
\end{figure*}

Fig. \ref{fig:7} compares the NMSE performance in CSI acquisition achieved by different schemes versus the number of pilot OFDM symbols $Q$. It can be observed that conventional CSI estimation schemes (e.g., GMMV-SOMP, GMMV-BSOMP, GMMV-AMP, and GMMV-LAMP algorithms) with csiNet for feedback cannot achieve a satisfactory sparse recovery performance with insufficient pilot and feedback signaling overhead, i.e., $Q < 50,\ B=32$. In contrast, the proposed CAN is capable of achieving an NMSE performance below 0.1 even when the number of pilot OFDM symbols is $Q=1$. 
Besides, even when compared to the conventional CSI estimation schemes with perfect CSI feedback, the proposed CAN still achieves a lower NMSE performance for $Q<50$.
Furthermore, the proposed CAN can achieve a lower NMSE for $Q\ge 2$ even when compared to the csiNet with perfect CSI estimation, which demonstrates the better CSI reconstruction performance of the Transformer architecture adopted by the proposed CAN compared to the CNN architecture adopted by the csiNet. The numerical results in Fig. \ref{fig:7} show that the proposed Transformer-based CAN is indeed a better CSI acquisition solution for RIS-aided Tera-Hertz massive MIMO systems with high-dimensional CSI but limited pilot and feedback signaling overhead.

Fig. \ref{fig:8} compares the NMSE performance achieved by different schemes versus the transmit power $P_t$.
It can be observed that both the conventional CS-based schemes and the model-driven DL-based GMMV-LAMP network fail to work with low transmit power (i.e., $P_t\le 10\ $dBm), while the proposed CAN still achieves good performance. 
Limited by the insufficient feedback signaling overhead $B=32$ and pilot signaling overhead $Q=16$, the NMSE performance of the proposed CAN will not further improve with the increase of the transmit power when $P_t\ge 30$ dBm.
The numerical results in Fig. \ref{fig:8} demonstrate the robustness of the proposed CAN for low transmit power.

\begin{figure*}[t]
	\vspace{-2mm}
	\captionsetup{font={footnotesize}, singlelinecheck = off, justification = raggedright,name={Fig.},labelsep=period}
	\centering
	\centering
	\hspace{-2mm}
	\begin{minipage}[t]{0.49\linewidth}
		\centering
		\includegraphics[scale=0.49]{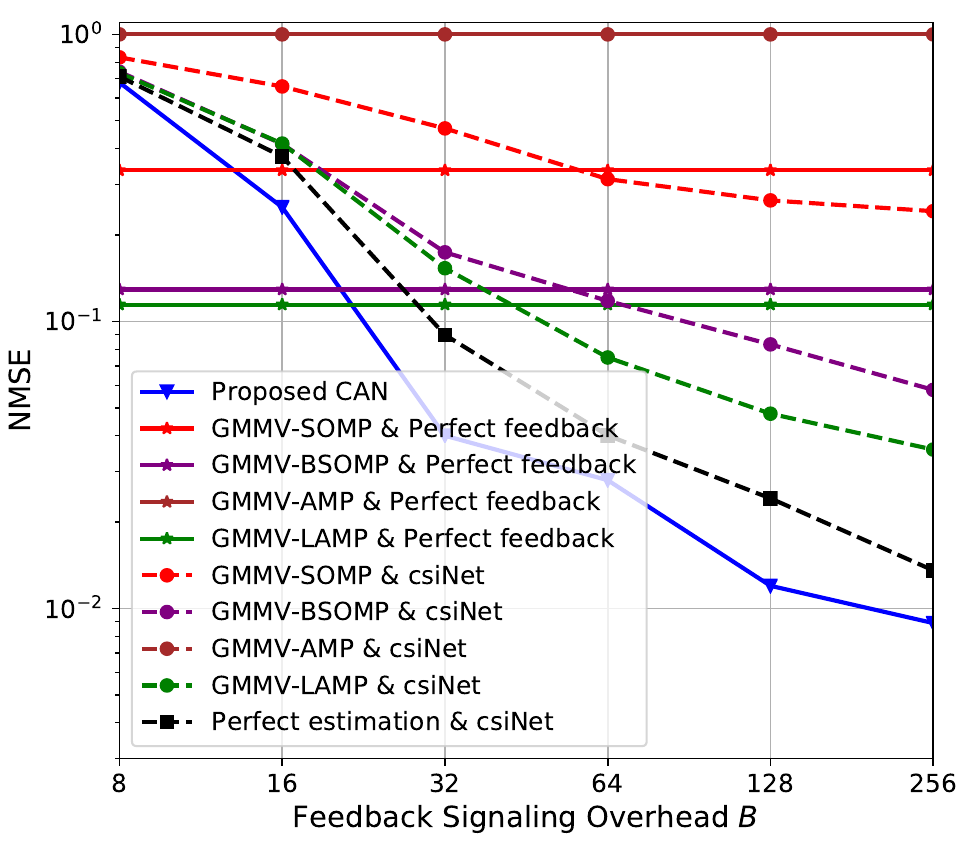}
		\vspace{-2.0mm}
		\captionsetup{font={footnotesize, color = {black}}, singlelinecheck = off, justification = raggedright,name={Fig.},labelsep=period}
		\caption{NMSE performance of different schemes versus the feedback signaling overhead $B$  ($Q=16$, $P_t=40\ {\rm dBm}$).}
		\label{fig:9}
	\end{minipage}
	\hfill
	\begin{minipage}[t]{0.49\linewidth}
		\centering
		\includegraphics[scale=0.49]{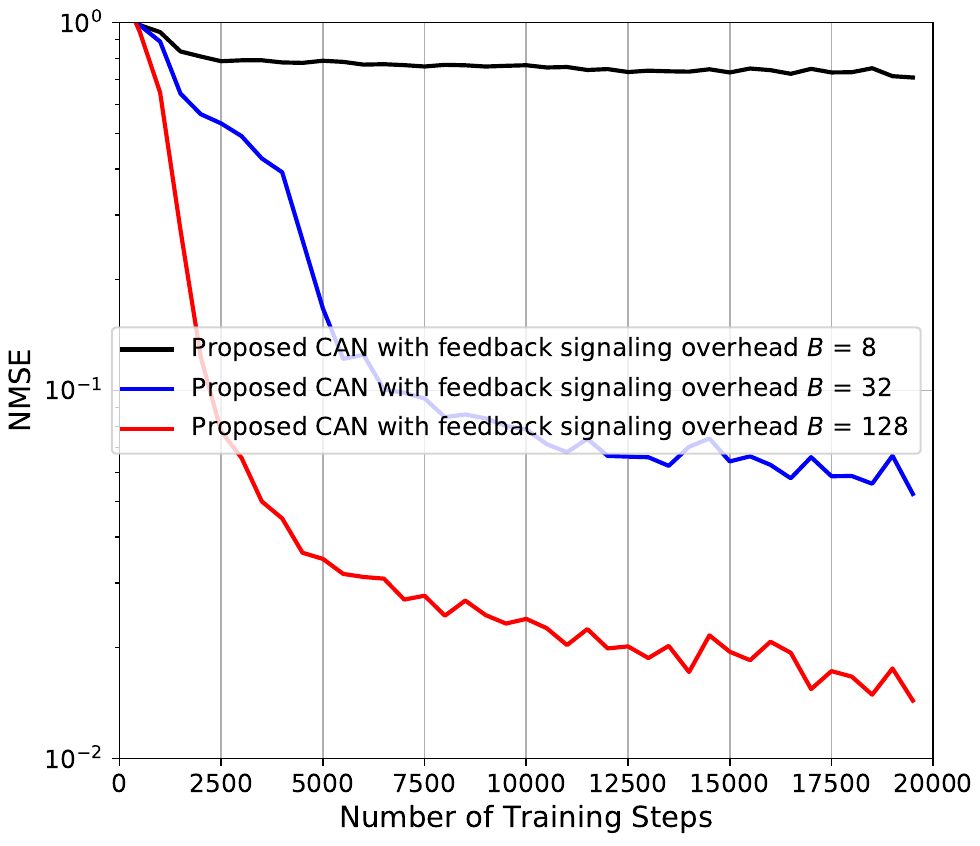}
		\vspace{-2.0mm}
		\caption{The convergence process of the proposed CAN ($Q=16$, $P_t=40\ {\rm dBm}$).}
		\label{fig:10}
	\end{minipage}
	
	\vspace{-2.0mm}
\end{figure*}

Fig. \ref{fig:9} shows the NMSE performance achieved by different schemes versus the feedback signaling overhead $B$. We can observe that the proposed CAN significantly outperforms the csiNet-based solutions. Furthermore, the proposed CAN requires only around 20 bits to achieve better NMSE performance than conventional CS and GMMV-LAMP schemes with perfect CSI feedback, which demonstrates the superior performance of the proposed CAN with limited feedback signaling overhead.

Fig. \ref{fig:10} compares the convergence process of the proposed CAN with different numbers of feedback bits. Note that in the simulation we use the warm-up strategy to adjust the learning rate according to formula (\ref{equ:warm}). It can be seen that the proposed CAN requires around 10,000 training steps (about 30 minutes of training time) to converge to the best NMSE performance.

\subsection{Performance Comparison of Different Precoding Schemes}

\begin{figure*}[t]
	\vspace{-6mm}
	\captionsetup{font={footnotesize}, singlelinecheck = off, justification = raggedright,name={Fig.},labelsep=period}
	\centering
	\centering
	\hspace{-2mm}
	\begin{minipage}[t]{0.49\linewidth}
		\centering
		\includegraphics[scale=0.49]{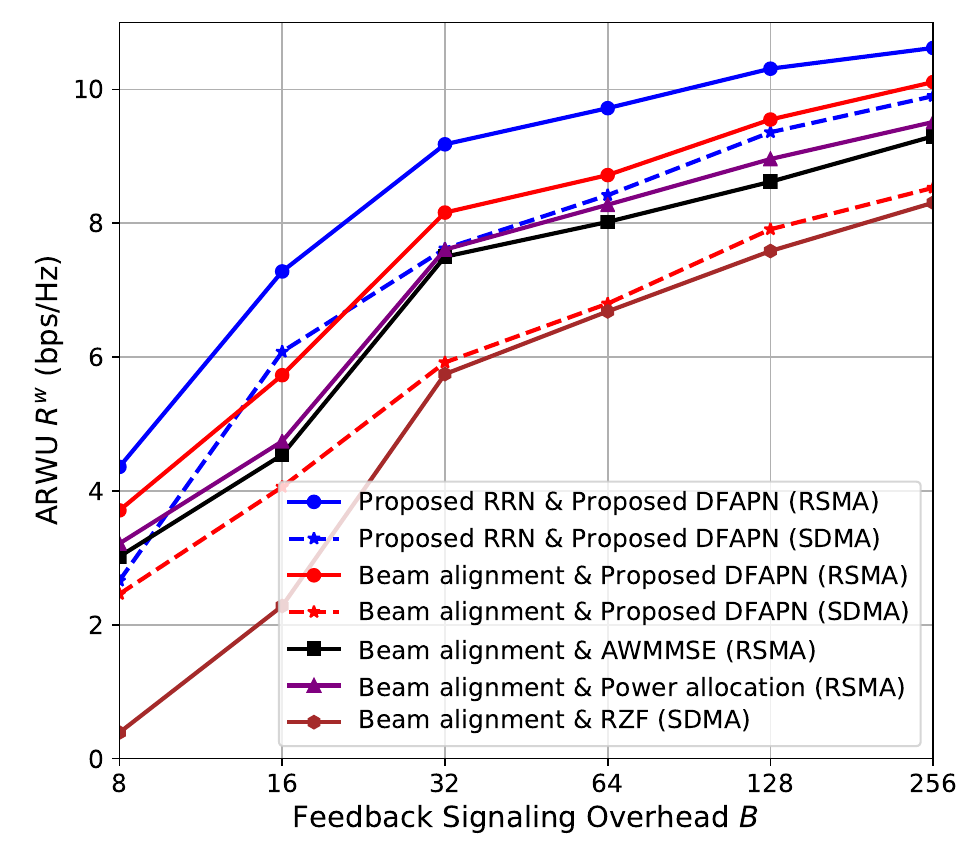}
		\vspace{-2.0mm}
		\captionsetup{font={footnotesize, color = {black}}, singlelinecheck = off, justification = raggedright,name={Fig.},labelsep=period}
		\caption{ARWU $R^w$ of different schemes  versus the feedback signaling overhead  $B$  ($Q=8$, $P_t=40\ {\rm dBm}$).}
		\label{fig:11}
	\end{minipage}
	\hfill
	\begin{minipage}[t]{0.49\linewidth}
		\centering
		\includegraphics[scale=0.49]{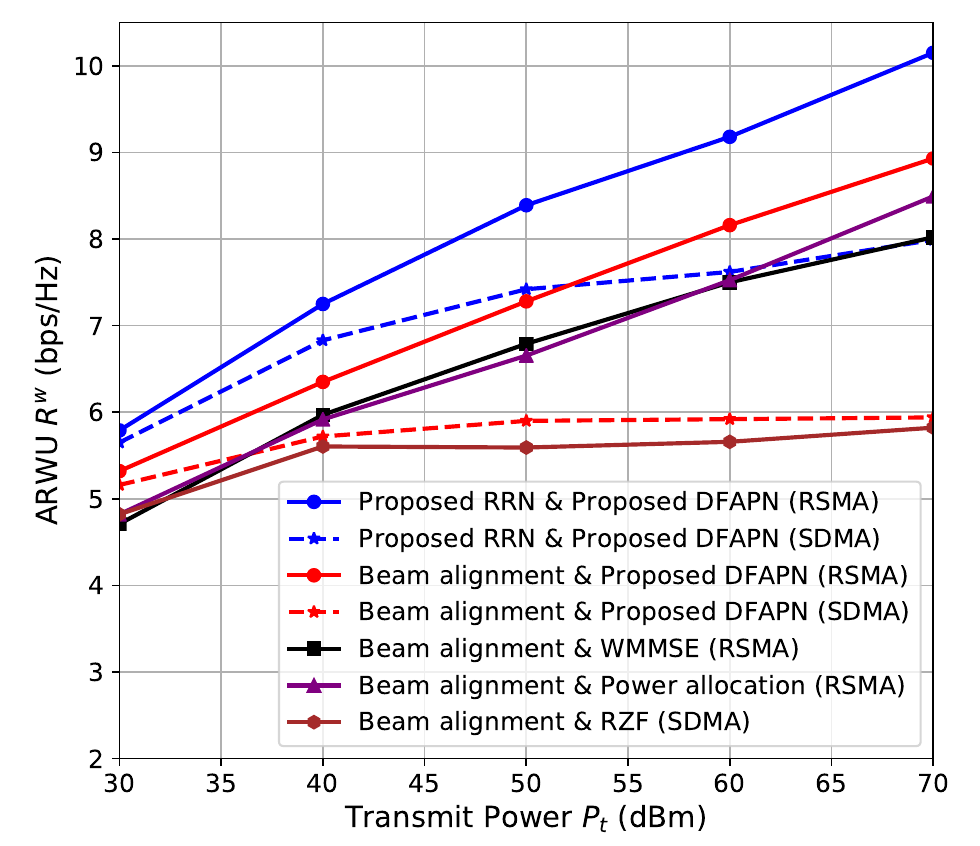}
		\vspace{-2.0mm}
		\caption{ARWU $R^w$ of different schemes  versus the transmit power  $P_t$  ($Q=8$, $B=32$).}
		\label{fig:12}
	\end{minipage}
	
	\vspace{-2.0mm}
\end{figure*}

Fig. \ref{fig:11} compares the ARWU performance achieved by different schemes versus the feedback signaling overhead  $B$.
It can be seen that the traditional RZF-based SDMA precoding scheme suffers from a severe performance loss when accurate CSI is not available at the BS due to the insufficient pilot and feedback signaling overhead. In contrast, by splitting the data streams into the common and private parts, the power allocation-based and AWMMSE-based RSMA precoding schemes achieve better performance even with very low pilot and feedback signaling overhead. The proposed DFAPN, however, is able to achieve better performance than the conventional schemes in both SDMA and RSMA scenarios by combining the AWMMSE model with DL.
This observation demonstrates that the model-driven DL-based scheme that combines a model-based approach with DL can significantly improve performance.
 Furthermore, it can be seen that adopting the proposed RRN (instead of conventional beam alignment-based RIS reflecting matrix design)  and jointly training the proposed RRN and DFAPN can achieve the best ARWU performance, which demonstrates the enhancement of the proposed RRN for the RIS reflecting matrix design. The numerical results in Fig. \ref{fig:11} demonstrate the superiority of the proposed RRN and DFAPN over the conventional schemes and the robustness of the proposed schemes to imperfect CSI with very low pilot and feedback signaling overhead.

Fig. \ref{fig:12} compares the ARWU performance achieved by different schemes versus the transmit power $P_t$.
It can be seen that the ARWU performance of the RSMA precoding schemes and the SDMA precoding schemes is similar at low transmit power regime (i.e., $P_t\le 20$ dBm). However, at high transmit power regime, the ARWU performance of the SDMA precoding schemes is limited by the inter-user interference caused by the CSI errors. In contrast, RSMA can significantly improve the robustness to imperfect CSI by splitting the data streams into common and private parts.
Furthermore, the proposed DL-based schemes has the best ARWU performance.
These observation demonstrate the effectiveness of the proposed  schemes in exploiting the available power for improving the system performance.

%

\begin{figure*}[t]
	\vspace{-2mm}
	\captionsetup{font={footnotesize}, singlelinecheck = off, justification = raggedright,name={Fig.},labelsep=period}
	\centering
	\centering
	\hspace{-2mm}
	
	\begin{minipage}[t]{0.49\linewidth}
		\centering
		\includegraphics[scale=0.49]{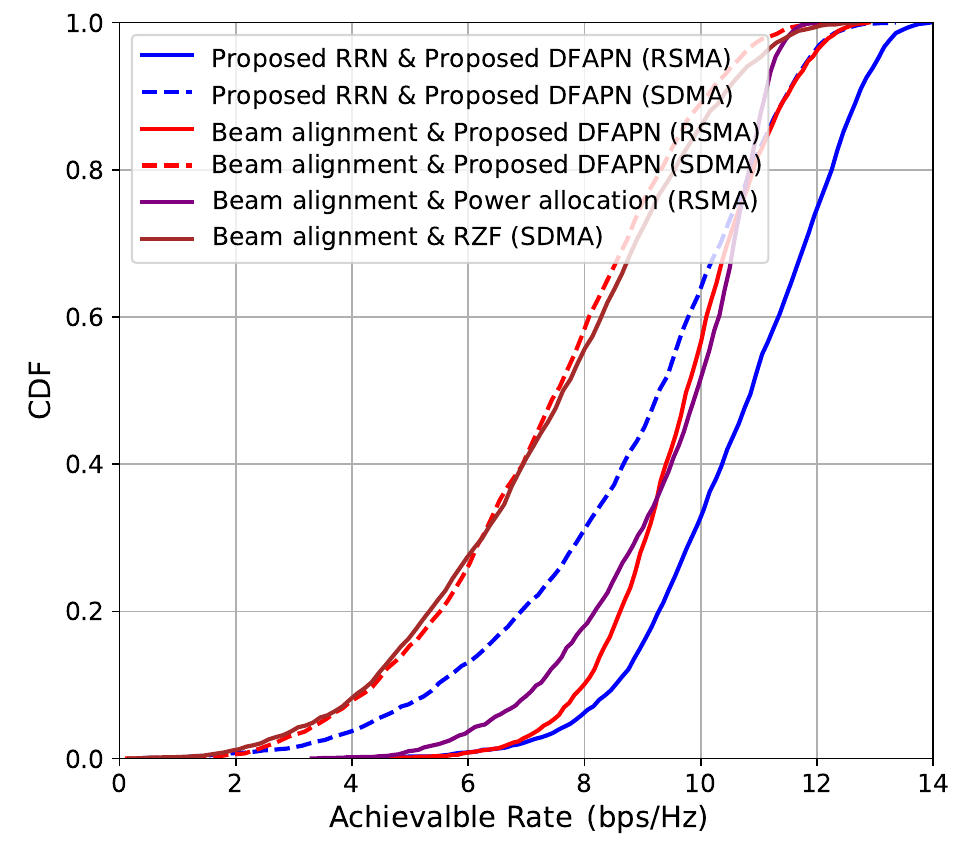}
		\vspace{-2.0mm}
		\caption{The CDF that describes the achievable rate of each UE achieved by different schemes  ($Q=8$, $B=32$, $P_t=50\ {\rm dBm}$).}
		\label{fig:13}
	\end{minipage}
	\hfill
	\begin{minipage}[t]{0.49\linewidth}
		\centering
		\includegraphics[scale=0.49]{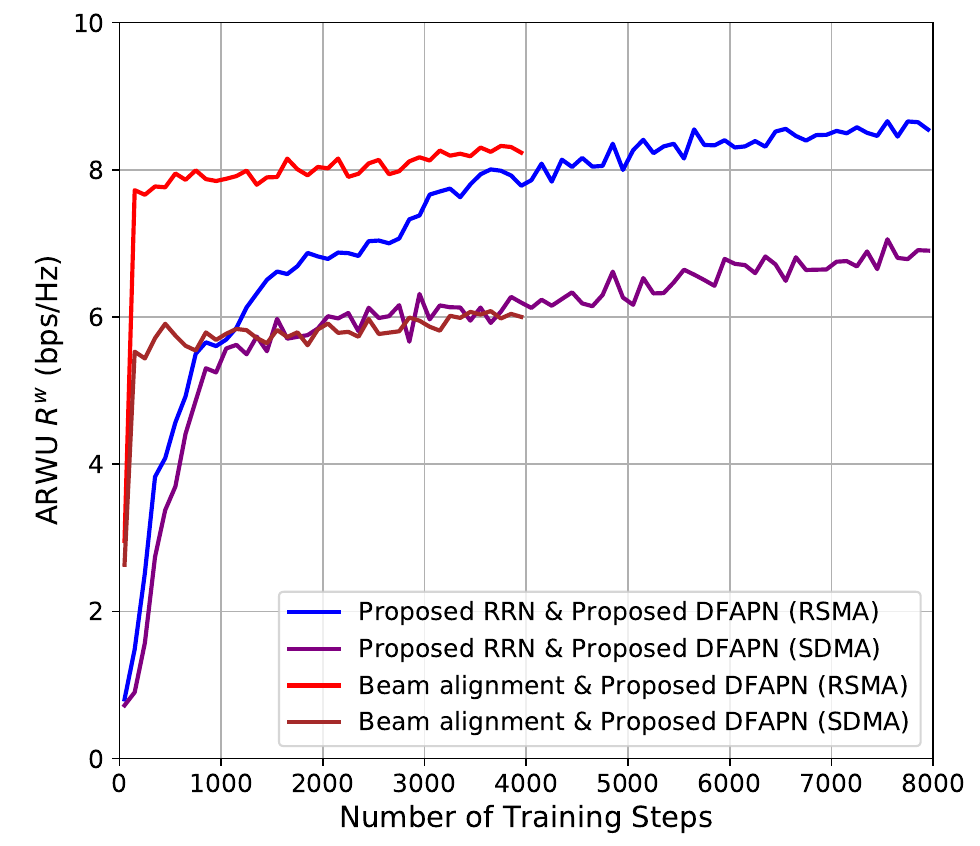}
		\vspace{-2.0mm}
		\captionsetup{font={footnotesize, color = {black}}, singlelinecheck = off, justification = raggedright,name={Fig.},labelsep=period}
		\caption{The convergence process of the proposed precoding shemes ($Q=8$, $B=32$, $P_t=40\ {\rm dBm}$).}
		\label{fig:14}
	\end{minipage}
	

	\vspace{-2.0mm}
\end{figure*}

\begin{figure*}[t]
	\vspace{-6mm}
	\captionsetup{font={footnotesize}, singlelinecheck = off, justification = raggedright,name={Fig.},labelsep=period}
	\centering
	\centering
	\hspace{-2mm}
	\begin{minipage}[t]{0.49\linewidth}
		\centering
		\includegraphics[scale=0.49]{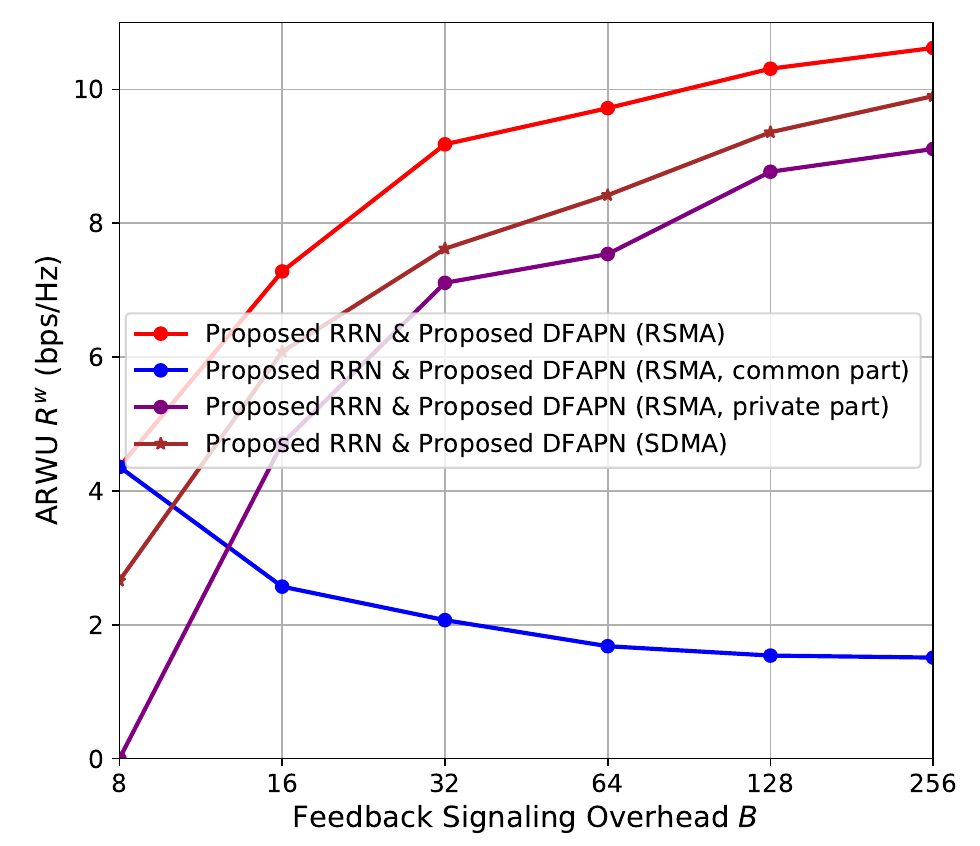}
		\vspace{-2.0mm}
		\captionsetup{font={footnotesize, color = {black}}, singlelinecheck = off, justification = raggedright,name={Fig.},labelsep=period}
		\caption{The different parts of the ARWU versus the feedback signaling overhead  $B$ for the proposed precoding schemes ($Q=8$, $P_t=40\ {\rm dBm}$).}
		\label{fig:15}
	\end{minipage}
	\hfill
	\begin{minipage}[t]{0.49\linewidth}
		\centering
		\includegraphics[scale=0.49]{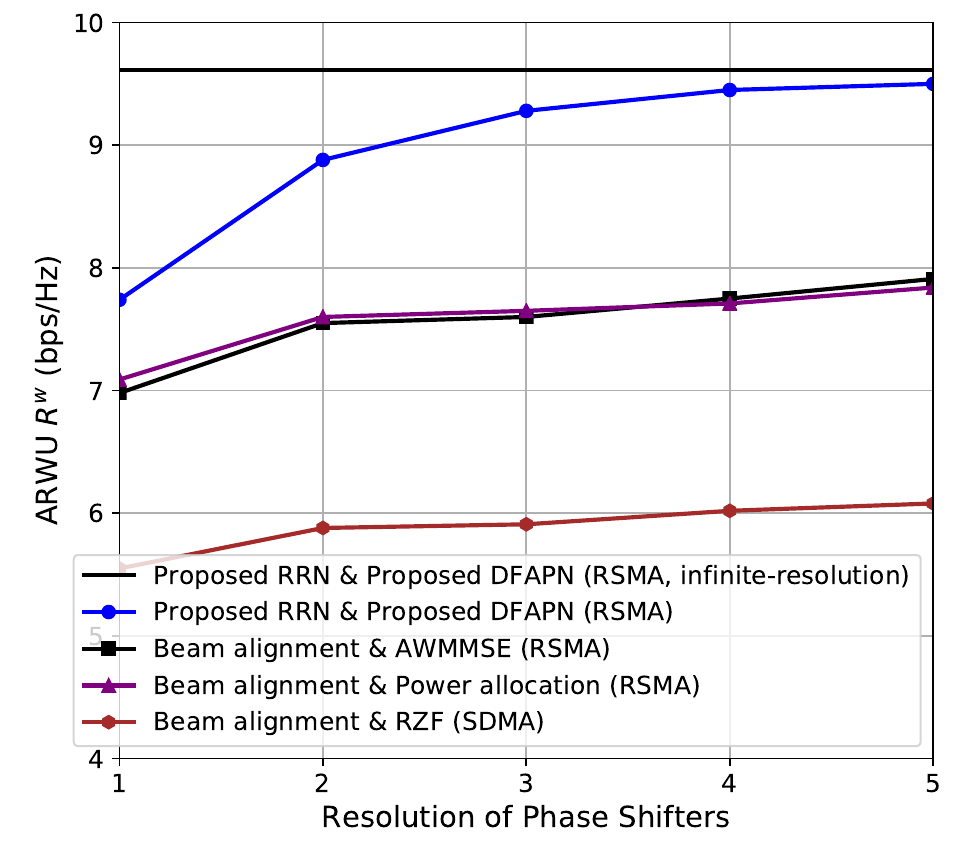}
		\vspace{-2.0mm}
		\captionsetup{font={footnotesize, color = {black}}, singlelinecheck = off, justification = raggedright,name={Fig.},labelsep=period}
		\caption{ARWU $R^w$ of the diferent precoding schemes versus the resolution of phase shifters ($Q=8$, $B=64$, $P_t=40$ dBm).}
		\label{fig:16}
	\end{minipage}
	
	\vspace{-2.0mm}
\end{figure*}

Fig. \ref{fig:13} shows the cumulative distribution function (CDF) that describes the achievable rate of each UE achieved by different schemes. 
Different from the previous simulations, here we consider observing the achievable rate performance of each UE rather than the ARWU performance, so as to clearly show the overall performance and the fairness between the UEs.
It can be observed that when the proposed DL-based RRN and DFAPN are adopted with the consideration of RSMA, about 95 percent of the UEs can achieve a rate of more than 8 bps/Hz, which is significantly better than the conventional schemes. Besides, the performance of the proposed DL-based RSMA precoding scheme can stably outperforman the conventional schemes on the CDF curve.
These observations show that, with high probability, the performance of the proposed DL-based RSMA precoding scheme will be good irrespective of RIS-UE channels.

%
%
%

Fig. \ref{fig:14} compares the convergence process of the proposed precoding schemes. It can be observed that when only the proposed DFAPN is trained for the digital active precoding and the conventional beam alignment-based RIS reflecting matrix design is used, only about 200 training steps (about 2 minutes of training time) are required for the proposed DFAPN to converge to the best ARWU performance. This observation demonstrates the advantage of the fast convergence of model-driven DL. In contrast, the joint training of the proposed RRN and DFAPN requires about 6000 training steps (about 1 hour of training time) to converge to the best ARWU performance, but can achieve better performance. We can select one of these schemes depending on our practical needs.

\begin{figure*}[t]
	\vspace{-2mm}
	\captionsetup{font={footnotesize}, singlelinecheck = off, justification = raggedright,name={Fig.},labelsep=period}
	\centering
	\centering
	\hspace{-2mm}
	\begin{minipage}[t]{0.49\linewidth}
		\centering
		\includegraphics[scale=0.49]{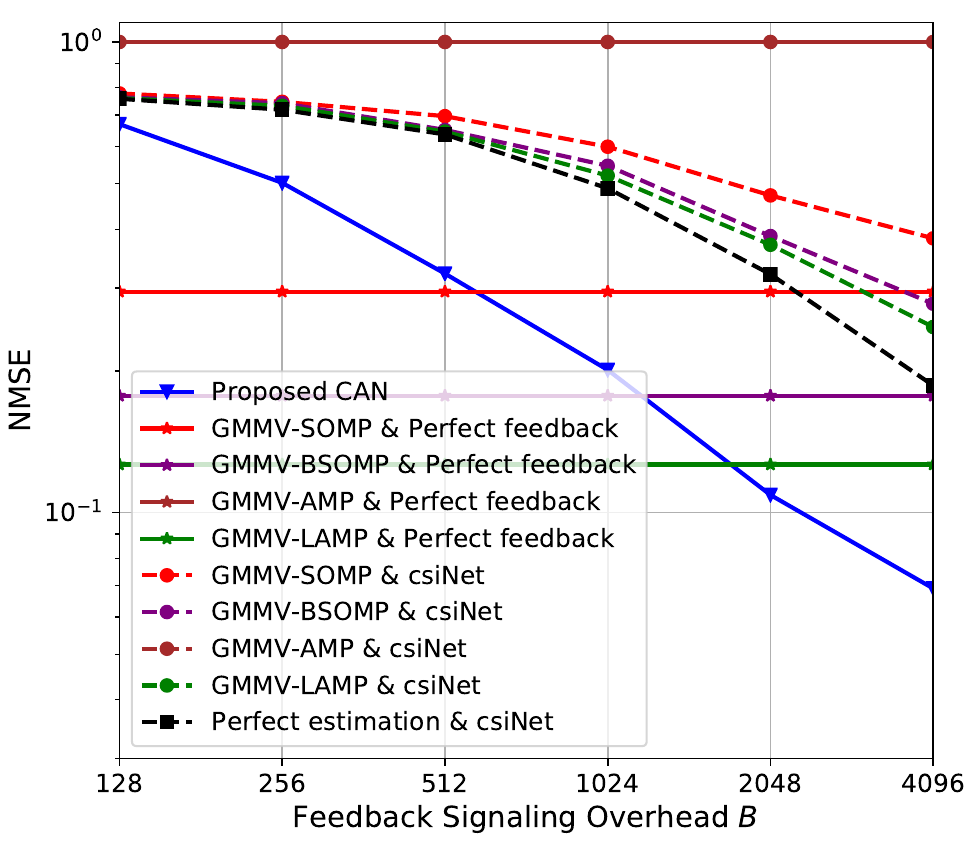}
		\vspace{-2.0mm}
		\captionsetup{font={footnotesize, color = {black}}, singlelinecheck = off, justification = raggedright,name={Fig.},labelsep=period}
		\caption{NMSE performance of different schemes versus the feedback signaling overhead $B$ in multipath scenarios  ($Q=32$, $P_t=40\ {\rm dBm}$).}
		\label{fig:17}
	\end{minipage}
	\hfill
	\begin{minipage}[t]{0.49\linewidth}
		\centering
		\includegraphics[scale=0.49]{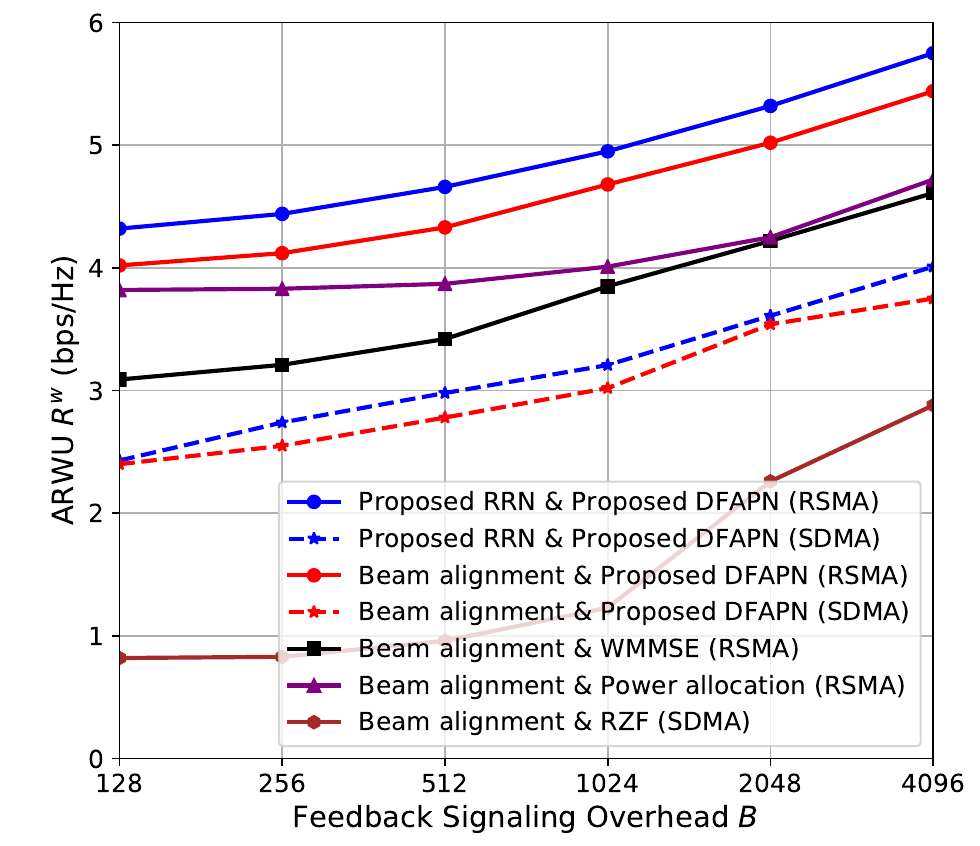}
		\vspace{-2.0mm}
		\caption{ARWU $R^w$ of different schemes  versus the feedback signaling overhead  $B$ in multipath scenarios ($Q=32$, $P_t=40\ {\rm dBm}$).}
		\label{fig:18}
	\end{minipage}
	
	\vspace{-2.0mm}
\end{figure*}

\begin{figure*}[t]
	\vspace{-6mm}
	\captionsetup{font={footnotesize}, singlelinecheck = off, justification = raggedright,name={Fig.},labelsep=period}
	\centering
	\centering
	\hspace{-2mm}
	\begin{minipage}[t]{0.49\linewidth}
		\centering
		\includegraphics[scale=0.49]{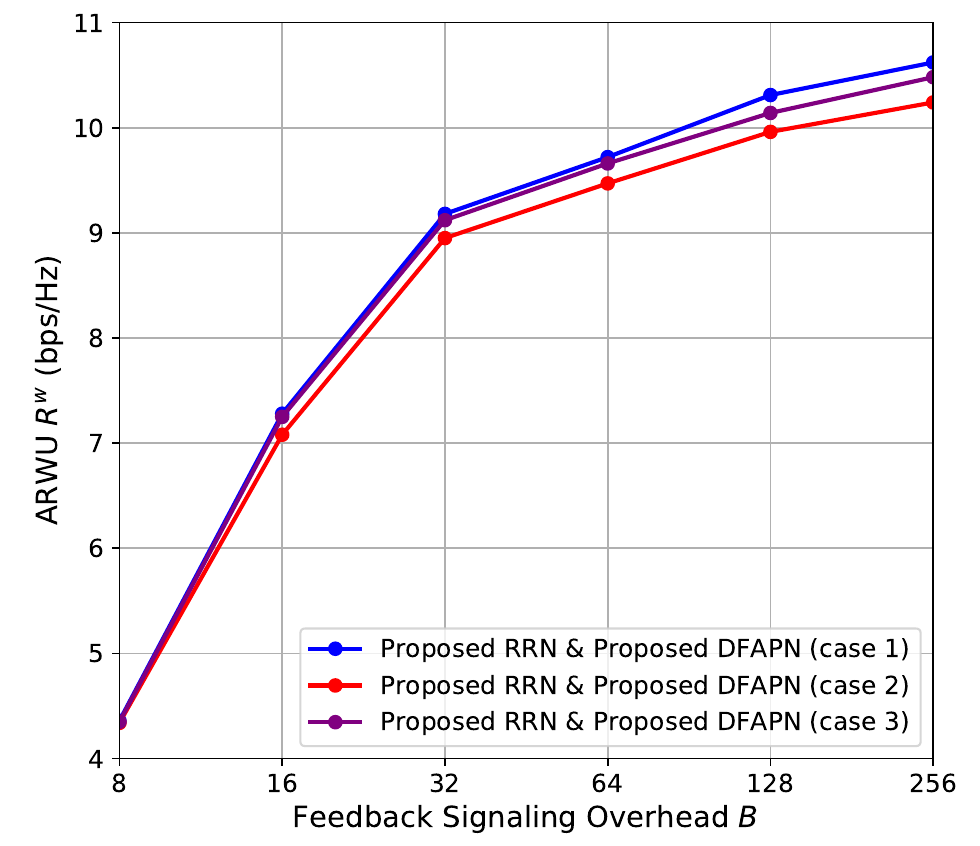}
		\vspace{-2.0mm}
		\captionsetup{font={footnotesize, color = {black}}, singlelinecheck = off, justification = raggedright,name={Fig.},labelsep=period}
		\caption{ARWU $R^w$ of the proposed precoding schemes versus the feedback signaling overhead  $B$  ($Q=8$, $P_t=40\ {\rm dBm}$).}
		\label{fig:19}
	\end{minipage}
	\hfill
	\begin{minipage}[t]{0.49\linewidth}
		\centering
		\includegraphics[scale=0.49]{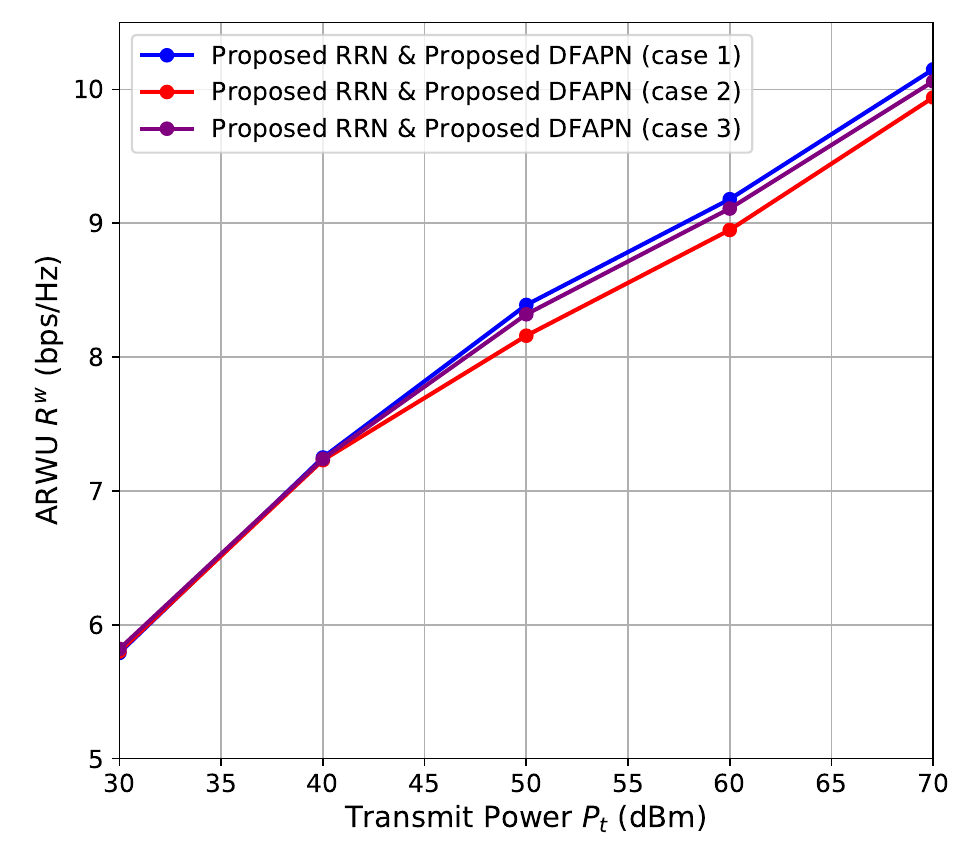}
		\vspace{-2.0mm}
		\captionsetup{font={footnotesize, color = {black}}, singlelinecheck = off, justification = raggedright,name={Fig.},labelsep=period}
		\caption{ARWU $R^w$ of the proposed precoding schemes  versus the transmit power  $P_t$  ($Q=8$, $B=32$).}
		\label{fig:20}
	\end{minipage}
	
	\vspace{-2.0mm}
\end{figure*}

{\color{black}Fig. \ref{fig:15} shows the different parts of the ARWU versus the feedback signaling overhead  $B$ for the proposed precoding schemes. It can be seen that when the feedback signaling overhead is insufficient, the BS cannot obtain an accurate CSI and the spatial multiplexing gain is difficult to exploit, thus a dominated portion of rate is allocated to the common messages by the proposed schemes. In contrast, as the feedback signaling overhead increases, the BS can obtain a more accurate CSI, and the spatial multiplexing gain can be better utilized, thus the proportion of private messages increases, exceeding that of the common messages.}

{\color{black}In the above simulations, we have considered RIS phase shifters with infinite-resolution, which is difficult to implement in practice. To this end, we further investigate the impact of the RIS phase shifter quantization error on the performance, as shown in Fig. \ref{fig:16}. It can be seen that the proposed RRN and DFAPN can still work even when RIS phase shifters have only 1 bit or 2 bits resolution.
Besides, the proposed RRN and DFAPN can achieve performance close to that with infinite-resolution phase shifters when the resolution of phase shifters is no less than 3 bits.}

\subsection{Performance Comparison in Multipath Scenario}

In the above simulations, we only considered RIS-UE channels with a single LoS path. To illustrate the generalizability of the proposed schemes, we further compare the performance of different schemes under the RIS-UE channels with multipath, as shown in Fig. \ref{fig:17} and Fig. \ref{fig:18}. In this case, the channel gain between the $j$-th reflecting element of the RIS and the $k$-th UE at the $n$-th subcarrier can be expressed as
\begin{equation} \label{equ:h_RU_multi}
	{h_{{\rm{RU}}}}[k,n,j] = \sum\limits_{l = 1}^{{L_p}} {{\beta ^l}[k]\sqrt {G_{{\rm{RU}}}^l[k,n,j]} \exp (\frac{{ - {\rm{j}}2\pi (d_{{\rm{RU}}}^l[j,k])}}{{{\lambda _n}}})},
\end{equation}
where $d_{{\rm{RU}}}^l[j,k]$ is the distance between the $j$-th reflecting element of the RIS and the $l$-th scatterer of the $k$-th UE, $G_{{\rm{RU}}}^l[k,n,j]$ is the corresponding large-scale fading gain, ${\beta ^l}[k] \sim {\cal CN}\left( {0},1 \right)$ is the path gain between the $k$-th UE and its $l$-th scatterer\footnote{The path gain from the scatterer to the UE, ${\beta ^l}[k]$, is related to the factors such as the cross-sectional area and the absorptivity of the scatterer. To simplify the mathematical expression, we directly model ${\beta ^l}[k]$ as a complex Gaussian random variable.}, and $L_p=4$ is the number of paths.

As shown in Fig. \ref{fig:17} and Fig. \ref{fig:18}, all the schemes require higher feedback signaling overhead than that in the single LoS path scenario due to the more complicated channels.
As for CSI acquisition, the proposed CAN can still acquire more accurate CSI than traditional schemes with lower feedback signaling overhead in this case. As for precoding, the proposed RRN and DFAPN are also able to perform RSMA precoding with imperfect CSI significantly better than other conventional schemes.
The numerical results in Fig. \ref{fig:17} and Fig. \ref{fig:18} demonstrate the generalizability of the proposed DL-based schemes to multipath channels.

{\color{black}
\subsection{The Impact of Ignoring the BS-UE Link}

In this paper, we consider that the LoS path of the BS-UE link is obscured, therefore the BS-UE link is weak. This is reasonable since the NLoS link is fairly weak in high-frequency communications,  espeically for Tera-Hertz systems \cite{Wan_Tcom}. Besides, the RSMA system is robust to CSI errors, since the utilized DL can adapt the proposed algorithm to practical data samples via training. Thus, we consider that ignoring the CSI of the BS-UE link has little impact on the performance of the system. To more rigorously justify the rationality of neglecting the BS-UE link, we add an additional set of simulations as shown in Fig. \ref{fig:19} and Fig. \ref{fig:20} to observe the impact of ignoring the BS-UE direct link. Specifically, we consider the following three simulation settings.

\begin{itemize}
	\item Case 1: The BS-UE link is ignored in both training and test stages of the proposed schemes.
	\item Case 2: The BS-UE link is ignored in the training stage of the proposed schemes, but considered in the test stage.
	\item Case 3: The BS-UE link is considered in both training and test stages of the proposed schemes.
\end{itemize}

Note that all the above cases are based on the proposed RRN and DFAPN in the RSMA scenario, and the CSI of the BS-UE link is generated in the same way as formula (46), where the path gain $\beta^l[k]$ is set to 0.01. As shown in Fig. \ref{fig:19} and Fig. \ref{fig:20}, ignoring the BS-UE link results in only a slight performance loss. Therefore, we believe that it is reasonable to ignore the BS-UE link to simplify the system model.}

\subsection{Performance Comparison in the Scenario with Larger Number of Array Elements}
\begin{figure*}[t]
	\vspace{-6mm}
	\captionsetup{font={footnotesize}, singlelinecheck = off, justification = raggedright,name={Fig.},labelsep=period}
	\centering
	\centering
	\hspace{-2mm}
	\begin{minipage}[t]{0.49\linewidth}
		\centering
		\includegraphics[scale=0.49]{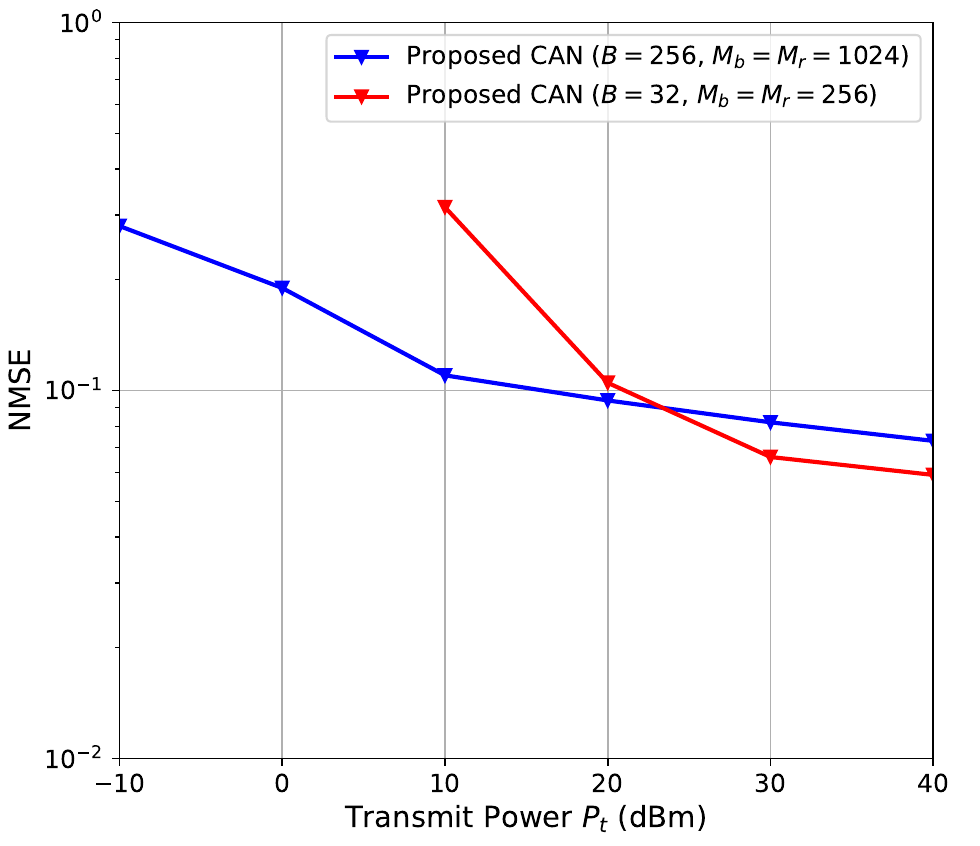}
		\vspace{-2.0mm}
		\captionsetup{font={footnotesize, color = {black}}, singlelinecheck = off, justification = raggedright,name={Fig.},labelsep=period}
		\caption{NMSE performance of different schemes versus the transmit power $P_t$, where $B = 256$, $Q = 8$, and both $M_b = M_r = 1024$ and $M_b = M_r = 256$ are provided.}
		\label{fig:21}
	\end{minipage}
	\hfill
	\begin{minipage}[t]{0.49\linewidth}
		\centering
		\includegraphics[scale=0.49]{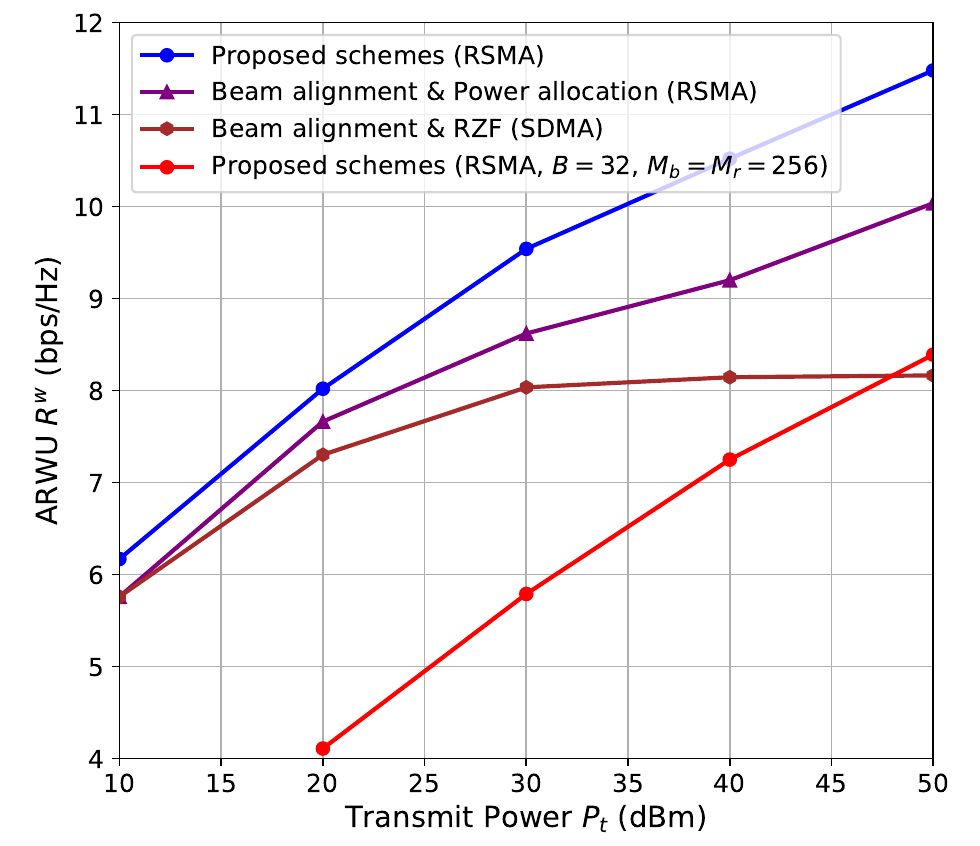}
		\vspace{-2.0mm}
		\captionsetup{font={footnotesize, color = {black}}, singlelinecheck = off, justification = raggedright,name={Fig.},labelsep=period}
		\caption{ARWU $R^w$ of different precoding schemes versus the transmit power  $P_t$  ($Q=8$, $B=256$, $M_b = M_r = 1024$). Here we also provide the proposed schemes (RSMA, $B=32$, $M_b=M_r=256$) for comparison.}
		\label{fig:22}
	\end{minipage}
	\vspace{-2mm}
\end{figure*}

To verify that the proposed schemes can be applied to larger RIS, we performed simulations in a scenario where the number of BS and RIS array elements is 1024.
The simulation results are shown in Fig. \ref{fig:21} and Fig. \ref{fig:22}, where the feedback overhead is $B=256$ and the pilot overhead is $Q=8$. Note that since the conventional CS-based channel estimation algorithms fail to reconstruct the CSI in the case of compression ratios up to 1024/8=128 (i.e., compress the signal from 1024-length to 8-length), we only show the CSI reconstruction results of the proposed CAN in Fig. \ref{fig:21}.
As can be seen from Fig. \ref{fig:21} and Fig. \ref{fig:22}, the proposed schemes are able to achieve good performance at larger RIS. Besides, due to the increase in the number of RIS and BS array elements, the beamforming gain is also larger, and the required transmit power can become smaller compared to the scenario with 256 array elements. Note that due to the increase of the CSI dimension, the feedback overhead in the scenario with 1024 array elements increases.

\subsection{Computational Complexity Analysis }\label{S5.6}
	
This subsection investigates the computational complexity of different schemes.
For the DL-based schemes, since there is no strict time limit at the offline training stage, we only consider the computational complexity at the online test stage. {\color{black}The computational complexity analysis of different schemes are presented in Table~\ref{table1}.
The details are as follows.}
\begin{itemize}
	\item The  GMMV-SOMP \cite{GMMV_SOMP}, GMMV-BSOMP \cite{GMMV_BSOMP}, GMMV-AMP \cite{GMMV_AMP}, or GMMV-LAMP \cite{Ma_JSAC} channel estimation algorithms share similar computational complexities, i.e., $\mathcal{O}\left(K Q G^{2} N_{c} I\right)$, which mainly comes from  matrix multiplication operations, where $G$ is the number of columns of the redundant dictionary matrix and $I$ is the number of iterations.
\end{itemize}
\begin{itemize}
	\item The computational complexity of the csiNet \cite{csinet1} mainly comes from $N_{\rm co}$ convolutional layers, i.e., $\mathcal{O}\left(\beta K U N_{c} \sum_{i=1}^{N_{c o}} n_{i-1} n_{i}\right)$, where $\beta$ is the size of the convolutional filters, $n_{i-1}$ and $n_i$ are the numbers of input and output feature maps of the $i$-th convolutional layer, respectively.
\end{itemize}
\begin{itemize}
	\item As for the proposed AWMMSE algorithm, we denote the number of iterations as $I_1$. In each iteration of the proposed AWMMSE algorithm, we adopt CVX toolbox to search and update the RSMA digital precoder. Assuming that the number of searches in each iteration is $I_2$, then the total computational complexity is $\mathcal{O}\left( K^2 N_{c} I_{1} I_{2}\right)$.
\end{itemize}
\begin{itemize}
	\item As for the proposed RRN, DFAPN, and CAN, the computational complexity mainly comes from the self-attention layers in Transformer, i.e.,  $\mathcal{O}\left(UN_c^2d_{\rm model}\right)$.
\end{itemize}

\begin{table*}[!tp]
	\vspace{-0mm}
	\centering  
	\color{black}
	\captionsetup{font={color = {black}}}
	\caption{Computational Complexity of Different Schemes.}  
	\label{table1}  
	\begin{tabular}{|c|l|}
		\hline
		\textbf{Schemes}                  & \multicolumn{1}{c|}{\textbf{Complexity}} \\ \hline
		GMMV-SOMP \cite{GMMV_SOMP}, GMMV-BSOMP  \cite{GMMV_BSOMP} & \multicolumn{1}{c|}{$\mathcal{O}\left(K Q G^{2} N_{c} I\right)$} \\ 
		GMMV-AMP \cite{GMMV_AMP} or GMMV-LAMP  \cite{Ma_JSAC}                    &           \\ \hline
		csiNet \cite{csinet1}                   & \multicolumn{1}{c|}{$\mathcal{O}\left(\beta KU N_{c} \sum_{i=1}^{N_{\rm c o}} n_{i-1} n_{i}\right)$}          \\ \hline
		Proposed AWMMSE                 & \multicolumn{1}{c|}{$\mathcal{O}\left( K^2 N_{c} I_{1} I_{2}\right)$}          \\ \hline
		Proposed CAN, RRN, or DFAPN & \multicolumn{1}{c|}{$\mathcal{O}\left(UN_c^2d_{\rm model}\right)$}         \\ \hline
		
	\end{tabular}
\vspace{-1mm}
\end{table*}

\begin{table*}[!tp]
	\centering  
	\color{black}
	\captionsetup{font={color = {black}}}
	\caption{Running Time of Different Schemes.}  
	\label{table2}  
	\begin{tabular}{|c|l|}
		\hline
		\textbf{Schemes}                  & \multicolumn{1}{c|}{\textbf{Running Time}} \\ \hline
		 GMMV-SOMP \cite{GMMV_SOMP},  & \multicolumn{1}{c|}{0.43-0.62 s in CPU} \\ 
		         GMMV-BSOMP  \cite{GMMV_BSOMP},     &           \\
		                or  GMMV-AMP \cite{GMMV_AMP}         &            \\
		  \hline
		  GMMV-LAMP  \cite{Ma_JSAC}                  & \multicolumn{1}{c|}{6.34 ms in GPU or 52.33 ms in CPU}          \\ \hline
		csiNet \cite{csinet1}                   & \multicolumn{1}{c|}{5.12 ms in GPU or 22.34 ms in CPU}          \\ \hline
		Proposed AWMMSE                 & \multicolumn{1}{c|}{70.14 s in CPU}         \\ \hline
		Proposed CAN, RRN, or DFAPN & \multicolumn{1}{c|}{3.24ms-5.63ms in GPU or 15.64ms-25.39ms in CPU}         \\ \hline
		
	\end{tabular}
	\vspace{-5mm}
\end{table*}

{\color{black}
To intuitively observe the computational complexity of different schemes, Table~\ref{table2} shows the running time of different schemes. It can be observed that the running time of DL-based schemes (i.e., \cite{csinet1,Ma_JSAC}, and the proposed DL-based schemes)  is significantly lower than that of model-based schemes (i.e., GMMV-SOMP \cite{GMMV_SOMP}, GMMV-BSOMP \cite{GMMV_BSOMP}, GMMV-AMP \cite{GMMV_AMP}, and AWMMSE).     }

\section{Conclusion}
This paper proposed a DL-based RSMA transmission solution for RIS-aided Tera-Hertz massive MIMO systems, where a robust RSMA precoding scheme and a CSI acquisition scheme with low pilot and feedback signaling overhead are conceived.
 Specifically, we first proposed a hybrid data-model driven DL-based RSMA precoding scheme, where the data-driven DL-based RRN was proposed for the RIS passive precoding, the model-based MF precoding strategy was proposed for the BS analog precoding, and the model-driven DL-based DFAPN was proposed for the BS active precoding.  
 As for the proposed DFAPN, we first derived a low-complexity AWMMSE for the digital RSMA active precoding at the BS, and further proposed the DFAPN by deep unfolding the proposed AWMMSE scheme for better precoding performance and lower computational complexity.
 We adopted the negative ARWU as the loss function to perform E2E training on the proposed RRN and DFAPN.
 Numerical results in the precoding showed that the proposed RRN and DFAPN are robust to imperfect CSI and have significantly better ARWU performance than the conventional schemes. 
Moreover, to obtain accurate CSI for better precoding performance,  we proposed the data-driven DL-based CAN to accurately acquire the downlink RIS-UE CSI at the BS with low pilot and feedback signaling overhead, where  the downlink pilot transmission, CSI feedback at the UE, and CSI reconstruction at the BS were modeled as an E2E neural network based on Transformer. 
We adopted NMSE as the loss function to perform E2E training on the proposed CAN, thereby improving the NMSE performance of the proposed CAN.
Numerical results in CSI acquisition showed that the proposed CAN can accurately estimate the downlink RIS-UE CSI at the BS with low pilot and feedback signaling overhead and low transmit power, while the conventional schemes suffer from a severv performance loss.


\end{document}